\documentclass[10pt,journal,compsoc]{IEEEtran}
\ifCLASSOPTIONcompsoc
  \usepackage[nocompress]{cite}
\else
  \usepackage{cite}
\fi
\ifCLASSINFOpdf
   \usepackage[pdftex]{graphicx}
\else
   \usepackage[dvips]{graphicx}
\fi
\usepackage{amsmath}
\usepackage{amsfonts}
\usepackage{array}
\usepackage{tabularx}
\usepackage{threeparttable}
\usepackage{multirow}
\usepackage{framed}
\usepackage{xcolor}
\usepackage{fancyhdr}
\usepackage[strings]{underscore}

\begin{document}
%
\title{SamBaS: \underline{Sam}pling-\underline{Ba}sed \underline{S}tochastic \\ Block Partitioning}
%
%
%

\author{Frank~Wanye, 
        Vitaliy~Gleyzer, 
        Edward~Kao,~\IEEEmembership{Member,~IEEE}
        and~Wu-chun~Feng,~\IEEEmembership{Senior~Member,~IEEE}
\thanks{DISTRIBUTION STATEMENT A. Approved for public release. Distribution is unlimited. This material is based upon work supported by the Under Secretary of Defense for Research and Engineering under Air Force Contract No. FA8702-15-D-0001. Any opinions, findings, conclusions or recommendations expressed in this material are those of the author(s) and do not necessarily reflect the views of the Under Secretary of Defense for Research and Engineering. \copyright 2022 Massachusetts Institute of Technology. Delivered to the U.S. Government with Unlimited Rights, as defined in DFARS Part 252.227-7013 or 7014 (Feb 2014). Notwithstanding any copyright notice, U.S. Government rights in this work are defined by DFARS 252.227-7013 or DFARS 252.227-7014 as detailed above. Use of this work other than as specifically authorized by the U.S. Government may violate any copyrights that exist in this work.

This project was supported in part by NSF I/UCRC CNS-1822080 via the NSF Center for Space, High-performance, and Resilient Computing (SHREC).

F. Wanye and W. Feng are with the Department of Computer Science, Virginia Tech, Blacksburg, VA 24060 USA (e-mail: wanyef@vt.edu; wfeng@vt.edu).

V. Gleyzer and E. Kao are with the MIT Lincoln Laboratory, Lexington, MA 02421 USA (e-mail: vgleyzer@mit.ll.edu; edward.kao@mit.ll.edu).}}

\IEEEtitleabstractindextext{%
\begin{abstract}

Community detection is a well-studied problem with applications in domains ranging from networking to bioinformatics. Due to the rapid growth in the volume of real-world data, there is growing interest in accelerating contemporary community detection algorithms. However, the more accurate and statistically robust methods tend to be hard to parallelize. 
One such method is stochastic block partitioning~(SBP) -- a community detection algorithm that works well on graphs with complex and heterogeneous community structure. In this paper, we present a \underline{sam}pling-\underline{ba}sed \underline{S}BP~(\textsf{SamBaS}) for accelerating SBP on sparse graphs. 
We characterize how various graph parameters affect the speedup and result quality of community detection with \textsf{SamBaS} and quantify the trade-offs therein. 
To evaluate \textsf{SamBas} on real-world web graphs without known ground-truth communities,
we
introduce \underline{p}artition \underline{q}uality \underline{s}core (PQS), an evaluation metric that outperforms modularity in terms of correlation with F1 score.
Overall, 
\textsf{SamBaS} achieves speedups of up to 10$\times$ while maintaining result quality (and even improving result quality by over 150\% on certain graphs, relative to F1 score).
\end{abstract}

}

\maketitle

\pagestyle{fancy}
\fancyhf{}
\fancyhead[L]{This work has been submitted to the IEEE for possible publication. Copyright may be transferred without notice, after which this version may no longer be accessible.}
\fancyfoot[R]{\thepage}
\thispagestyle{fancy}

\definecolor{darkorange}{rgb}{.60,.25,.00}
\newcommand{\wu}[1]{
  \begin{framed}
    {\noindent\color{darkorange}{\em #1 -- Wu}}
  \end{framed}
}

\definecolor{blue}{rgb}{.25,.25,.70}
\newcommand{\frank}[1]{
  \begin{framed}
    {\noindent\color{blue}{\em #1 -- Frank}}
  \end{framed}
}

\IEEEdisplaynontitleabstractindextext

%
\IEEEpeerreviewmaketitle

\IEEEraisesectionheading{\section{Introduction}\label{sec:introduction}}


\IEEEPARstart{D}{ATA} from a wide variety of sources, including domains such as bioinformatics~\cite{PereiraLeal2003DetectionNetworks}, the world wide web~\cite{Huberman1999GrowthWeb}, and computer networks~\cite{Krishnamurthy2000OnClients}, can be represented as graphs. In this representation,
graph vertices represent individual entities, and graph edges represent the relationships between them.
These edges can be directed or undirected as well as weighted or unweighted. 

In real-world graph data, it is often possible to group vertices into communities such that vertices within a community are more strongly connected to each other than they are to vertices outside their community. These communities can either be non-overlapping, where a single vertex belongs to at most one community, or overlapping, where a single vertex belongs to more than one community. This grouping process is known as \emph{community detection}.

Community detection has a multitude of applications in a variety of domains. In computational clusters, community detection can be used to reduce the communication overhead when assigning workloads to computational nodes~\cite{Levchuk2018OptimizingGraphs}. In biological networks, identifying communities of protein interactions can help with the identification of functional modules~\cite{PereiraLeal2003DetectionNetworks}. It has also been shown to improve the accuracy of classification tasks~\cite{Rizos2017MultilabelNetworks,Li2017PacketDetection} and to help inform the placement of servers in communication networks~\cite{Krishnamurthy2000OnClients}.

However, optimal community detection is an NP-hard problem~\cite{Fortunato2010CommunityGraphs}, making it intractable for real-world graphs, which can consist of billions of vertices and edges. As such, a number of sub-optimal approaches have been developed to facilitate community detection on large graphs. These approaches are based on heuristics such as minimum cut, vertex similarity, modularity maximization, and statistical inference~\cite{Fortunato2010CommunityGraphs}.

Of the aforementioned heuristics, \emph{modularity maximization}, where vertices are grouped into communities  to maximize the number of edges between vertices in the same communities, is a commonly applied heuristic~\cite{Kao2017StreamingPartition} due to its relative speed and scalability~\cite{Blondel2008FastNetworks}. The popularity of modularity maximization has also led to substantial work into parallel and distributed implementations of algorithms based on this heuristic~\cite{Que2015ScalableAlgorithm,Ghosh2019ScalingClustering,Ghosh2018DistributedDetection,Lu2015ParallelDetection,Ghosh2018ScalableVite}. However, modularity maximization has a resolution limit that limits the size of the communities that can be detected with these methods~\cite{Fortunato2007ResolutionDetection,Peixoto2014HierarchicalNetworks} and affects the quality of results achieved when they are applied to graphs where the community sizes vary widely.

\begin{figure}[ht!]
  \centering
  \includegraphics[width=\columnwidth]{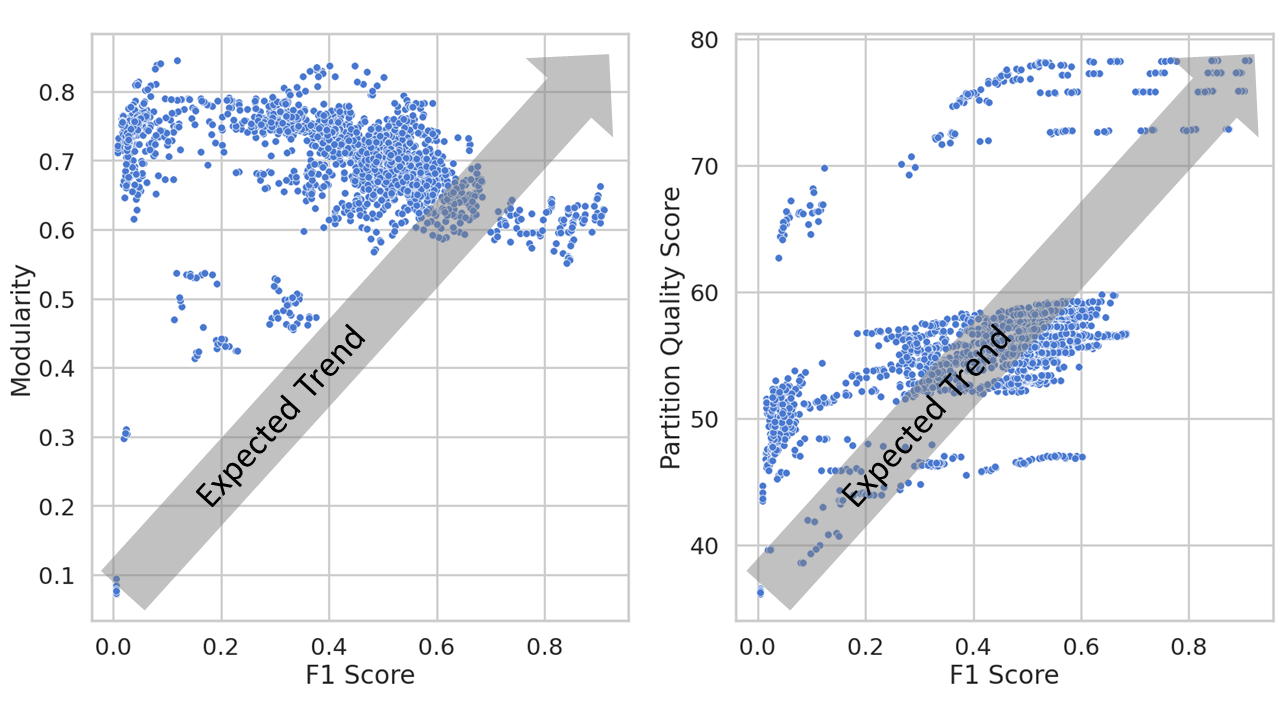}
  \caption{Comparison of the relationships between F1 Score and Modularity (left) and F1 Score and Partition Quality Score (right), with points from community detection runs on synthetic graphs across different sampling approaches and sample sizes. Partition Quality Score increases as community detection accuracy based on truth (F1 Score) increases; Modularity does not display this expected trend.}
  \label{fig:metrics}
\end{figure}


To address the above issues, we propose and design a data-sampling approach based on the stochastic block partitioning~(SBP) algorithm. SBP~\cite{Peixoto2013ParsimoniousNetworks,Peixoto2014EfficientModels} is a statistical inference-based algorithm for performing non-overlapping community detection on unattributed graphs. SBP has a relatively low algorithmic complexity of $O(|E|\log^2|E|)$ in terms of the number of edges $|E|$ in the graph. Furthermore, it does \textit{not} suffer from the resolution limit because it is \textit{not} based on the modularity metric~\cite{Kao2017StreamingPartition}. More importantly, SBP is robust on graphs where the community sizes vary widely and the connectivity between the communities is high, making it especially desirable for graphs where communities are hard to detect. However, it suffers from scalability issues, making its serial implementation infeasible for large graphs. Additionally, the algorithm is based on an inherently serial inference technique, making parallelization nontrivial~\cite{Kao2017StreamingPartition}.

Several attempts have been made to improve the runtime of the SBP algorithm, including via graph streaming~\cite{Uppal2018FastGraphs} and distributed processing on computational clusters~\cite{Uppal2017ScalablePartition}. The streaming approach seeks to speed up community detection by streaming the graph in batches. 
Our sampling approach, which was prototyped in~\cite{Wanye2019FastSampling}, complements the above by focusing on data reduction for accelerating SBP. Combining our approach with either of the above would lead to further speedup gains.

In this manuscript, we introduce \textbf{\underline{sam}}pling-\textbf{\underline{ba}}sed 
\textbf{\underline{S}}BP (SamBaS), a holistic sampling method for accelerating stochastic block partitioning (SBP). We characterize the accuracy and runtime performance of SamBaS under different graph parameters and show that SamBaS speeds up community detection without significantly affecting accuracy via analysis on 10 real-world graphs with no known ground truth. Whereas modularity is typically used in such scenarios to measure the quality of community detection results, our experiments on realistic synthetic graphs show that modularity does not correlate well with the F1 score. Thus, we introduce a new metric based on the \emph{minimum description length} 
to measure the quality of community detection results in graphs with no known truth.


\vspace*{6pt}
\noindent\textbf{Contributions.}
This manuscript makes the following contributions:
\begin{itemize}
    \item A holistic \underline{sam}pling-\underline{ba}sed approach for accelerating \underline{s}tochastic block partitioning (SamBaS) \emph{without} sacrificing accuracy.
    \item A rigorous characterization of the accuracy and computational performance of SamBaS 
    with varying graph parameters, compared to community detection~(SBP) without sampling.
    \item A novel metric, \textit{partition quality score} (PQS), for evaluating community detection in real-world graphs with no known ground-truth communities. We empirically show that PQS is a  better measure of community detection accuracy than modularity~(see Fig.~\ref{fig:metrics}).
\end{itemize}

Overall, we show that SamBaS can lead to speedups of up to 10$\times$ on synthetic graphs and up to 21$\times$ on real-world graphs while maintaining result quality at comparable levels to SBP without sampling. On some graphs, it can even improve result quality by upwards of 100\% (see Fig.~\ref{fig:intro}). We attribute this improvement in quality to the effect that search-space reduction via sampling has on SBP.

\section{Background and Related Work}


In this section, we introduce existing research related to community detection, accelerating community detection, and sampling from graphs.

\subsection{Stochastic Blockmodels}\label{sec:blockmodels}

Stochastic Blockmodels~(SBMs) are a set of graph models that can be used to generate and perform inference on graphs. They represent the graph as a matrix of probabilities of edges existing between communities. How well a blockmodel represents a graph $G$ can be calculated using the unnormalized log-likelihood of the graph given the model, $L(G|B)$. For the variant of blockmodel we use in this manuscript, the Degree-Corrected SBM~(DCSBM), the nonparametric $L(G|B)$ is calculated using the following equation~\cite{Peixoto2017NonparametricModel}:

\begin{equation}\label{eq:mdcsbm_log_likelihood}
    L(G|B) = \ln \left(
    \frac{\prod_{i,j}B_{i,j}!}{\prod_{i}e_{i,out}!\prod_{i}e_{i,in}!}
    \times
    \frac{\prod_{k}d_{k,out}!\prod_{k}d_{k,in}!}{\prod_{k,l}M_{k,l}!}
    \right),
\end{equation}

\begin{figure*}[ht!]
  \centering
  \includegraphics[width=0.6\textwidth]{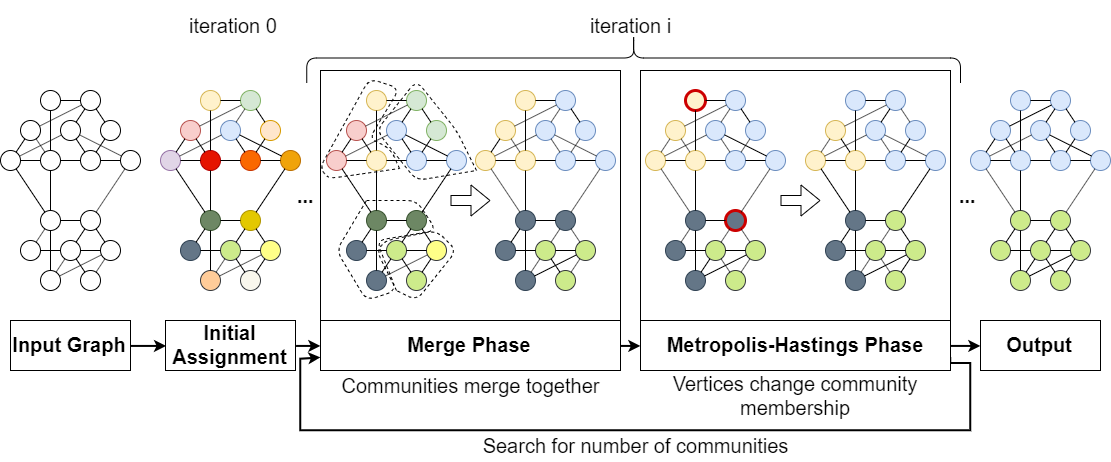}
  \caption{Snapshots of a graph at various stages of the stochastic block partitioning algorithm.}
  \label{fig:sbp}
\end{figure*}

\noindent
where $M_{k,l}$ is the adjacency matrix term corresponding to vertices $v_k$ and $v_l$, $B_{i,j}$ refers to the number of edges going from community $i$ to community $j$, $e_{i,in}$ and $e_{i,out}$ are the number of incoming and outgoing edges of vertices in community $i$, and $d_{k,out}$ and $d_{k,in}$ refer to the out- and in-degree of vertex $v_k$, respectively. For a derivation of Equation~\ref{eq:mdcsbm_log_likelihood}, we refer the reader to~\cite{Peixoto2017NonparametricModel}.

\subsection{Stochastic Block Partitioning}\label{sec:sbp}

The Stochastic Block Partitioning~(SBP)~\cite{Peixoto2013ParsimoniousNetworks,Peixoto2014EfficientModels,Kao2017StreamingPartition} algorithm is an iterative, agglomerative statistical inference method for community detection, based on inference over the Degree-Corrected Stochastic Blockmodel~(DCSBM)~\cite{Peixoto2017NonparametricModel}. This inference can be done by minimizing the entropy of the blockmodel, but because the blockmodel has the lowest entropy when each vertex is assigned a unique community, that only works when the number of communities is known beforehand. In order for the model to be able to discover the optimal number of communities, the SBP algorithm minimizes the minimal description length $H$ of the DCSBM, which represents the total amount of information needed to describe the blockmodel itself. Given a graph $G$ consisting of $|E|$ edges and $|V|$ vertices, described by a blockmodel with $|C|$ communities, $H$ is given by~\cite{Peixoto2017NonparametricModel}:
\begin{equation}\label{eq:mdl}
    H = -\ln(B) - L(B|G),
\end{equation}
\noindent where $L(B|G)$ is computed using Eq.~(\ref{eq:mdcsbm_log_likelihood}) and $\ln(B)$ is computed as follows:

\begin{align}
\begin{split}
    \ln(B) = & \sum\limits_{i}\ln|C_i|! - \ln\left(\binom{|C|(|C|+1)/2}{|E|}\right) \\
    & - \ln\binom{|V|-1}{|C|-1} - \ln|V|! \\
    & + \ln \left(
    \prod_{i}\frac{\prod_{d}|v_d^i|!}{|C_i|}\prod_{i}q(e_{i},|C_i|)^{-1}
    \right)
    ,
\end{split}
\end{align}

\noindent where 
$\left(\binom{n}{m}\right) = \binom{n + m + 1}{m}$, 
$\binom{n}{m} = \frac{n!}{m!(n-m)!}$, 
$|v_d^i|$ is the number of vertices in block $i$ with degree $d$, $|C_i|$ is the number of vertices in block $i$, and $q(n,m)$ refers to the ``number of restricted partitions of integer $a$ into at most $b$ parts''. For a more in-depth discussion of $H$, we refer the reader to~\cite{Peixoto2017NonparametricModel}.

Initially, each vertex is treated as a separate community. Each iteration of the algorithm consists of two phases. In the first phase, communities are merged together based on the resulting difference in $H$. In the second phase, the community memberships are fine-tuned at the vertex level using the Metropolis-Hastings algorithm~\cite{Hastings1970Monteapplications}. By keeping up to three copies of the DCSBMs obtained with different numbers of communities, a search can be performed to find the optimal number of communities and the optimal DCSBM for this number of communities. The algorithm is visually summarized in Figure~\ref{fig:sbp}.

Although this algorithm is slower and harder to parallelize than several alternatives, we focus on it because of its robustness to (a) high variation in community sizes and (b) complex community structure as characterized by high intercommunity connectivity in real-world graphs. This is corroborated by comparing the results obtained by SBP~\cite{Ghosh2019ScalingClustering}, Fast-Tracking Resistance~\cite{Ghosh2019ScalingClustering}, Louvain~\cite{Ghosh2019ScalingClustering,Ghosh2018ScalableVite}, and Label Propagation~\cite{Liu2019DistributedDetection} on the Graph Challenge~\cite{Kao2017StreamingPartition} datasets. Across all those implementations, SBP delivered the best and most consistent performance on graphs with complex and heterogeneous structure. Furthermore, 
running the \verb|graph-tool| SBP implementation~\cite{Peixoto2014TheLibrary} on the same graphs led to even better results than those reported in~\cite{Ghosh2019ScalingClustering}.

\subsection{Accelerating Community Detection}

Much of the research into accelerating community detection algorithms has centered around modularity maximization algorithms like Louvain and fast heuristics such as label propagation, which have been parallelized on multi-core~\cite{Halappanavar2017ScalableGrappolo,Lu2015ParallelDetection}, multi-node~\cite{Liu2019DistributedDetection,Ghosh2019ScalingClustering,Ghosh2018DistributedDetection,Que2015ScalableAlgorithm}, and heterogeneous architectures~\cite{Naim2017CommunityGPU}.

Speeding up the SBP algorithm is challenging and has not received as much research attention. 
The Metropolis-Hastings portion of SBP is an inherently serial Markov Chain Monte-Carlo~(MCMC) method: every time a vertex move is accepted, the DCSBM changes, which affects the probabilities of all subsequent vertex moves being accepted or rejected. Parallelizing such MCMC algorithms is an open problem, with different challenges arising depending on how the algorithm is parallelized~\cite{Wilkinson2005ParallelComputation}. In~\cite{Uppal2017ScalablePartition}, a parallel and distributed SBP implementation is proposed based on random vertex partitioning and a heuristic combination of multiple independent MCMC chains. However, it has several limitations, including possible bottlenecks due to the use of a single message aggregator thread, and a drop-off in result quality when the number of nodes across which the algorithm is distributed is high.

Our approach leverages an existing partially parallelized implementation of SBP~\cite{Peixoto2014TheLibrary} while focusing on delivering further speedup gains through data reduction.

\subsection{Data Reduction in Graphs}

Reducing the size of graphs before storing or processing is a common technique, including agglomerative approaches that combine multiple vertices into a single super-node~\cite{Stanley2018CompressingNodes}, selective edge deletion~\cite{Besta2019SlimGraph}, 
and various sampling techniques~\cite{Besta2019SlimGraph,Leskovec2006SamplingGraphs,Wang2011UnderstandingAnalysis,Maiya2010SamplingStructure}. In this work, we focus on vertex-sampling techniques for their relative simplicity and because removing a vertex from a graph also removes the edges connected to it, thus reducing the size of the graph along both the vertex and edge dimensions.

Graphs have a multitude of properties that describe the graph, including the graph diameter, vertex degree distribution, sparsity, strongly connected components, and clustering coefficient. A "perfect" data reduction scheme would preserve all of these properties. Unsurprisingly, this is an impossible task with sampling. Studies into sampling algorithms~\cite{Leskovec2006SamplingGraphs,Wang2011UnderstandingAnalysis,Besta2019SlimGraph,Maiya2010SamplingStructure} show that different sampling algorithms excel in preserving different subsets of a graph's properties, with no one algorithm consistently performing well across all properties.

Data reduction algorithms have been previously studied in the context of community detection (see Table~\ref{tab:related_work_summary} for a summary of such work). 
Gao and Maiya~\cite{Gao2016AcceleratingSampling,Maiya2010SamplingStructure}
separately showed that sampling can preserve community structure and that sampling can be used with several community detection algorithms without a significant loss in accuracy. Stanley~\cite{Stanley2018CompressingNodes} showed that agglomerative supernodes can be used to reduce the variability in community detection results from run to run.

However, none of these works perform an in-depth study into the effects of different graph properties on the viability of data reduction for community detection. 
Neither do they explain the variability in community detection results using data reduction on different graphs, as shown in our preliminary results~(see \cite{Wanye2019FastSampling}).
Finally, only~\cite{Stanley2018CompressingNodes} considers any statistics-based community detection algorithms, which may be more significantly affected by the data reduction approach, in their analysis.

\begin{table}[ht!]
\centering
\caption{Summary of Related Data Reduction Work.}
\label{tab:related_work_summary}
\resizebox{\columnwidth}{!}{%
\begin{threeparttable}
\begin{tabular}{c||c|c||c|c}
\hline
\multirow{3}{*}{Paper} & Community & Data & \multicolumn{2}{c}{Evaluation} \\
\cline{4-5}
 & Detection & Reduction & Graph & Runtime \\
 & Algorithm & Method & Properties\tnote{1} & Performance \\
\hline
Maiya~\cite{Maiya2010SamplingStructure} & GN, NLE, CNM\tnote{2} & Sampling & & \\
Gao~\cite{Gao2016AcceleratingSampling} & LP\tnote{2}, Louvain & Sampling & & \checkmark \\
Stanley~\cite{Stanley2018CompressingNodes} & SBP, Louvain & Supernodes & & \checkmark \\
\hline
This Manuscript & SBP & Sampling & \checkmark & \checkmark \\
\hline
\end{tabular}
\begin{tablenotes}
\item[1] Evaluates the effect of different graph properties on results obtained with the data reduction approach.
\item[2] Label Propagation (LP), Girvan-Newman (GN), Newman's Leading Eigenvector (NLE), Clauset-Newman-Moore (CNM).
\end{tablenotes}
\end{threeparttable}
} 
\end{table}

\section{Methodology}

\begin{figure*}[ht!]
  \centering
  \includegraphics[width=0.6\textwidth]{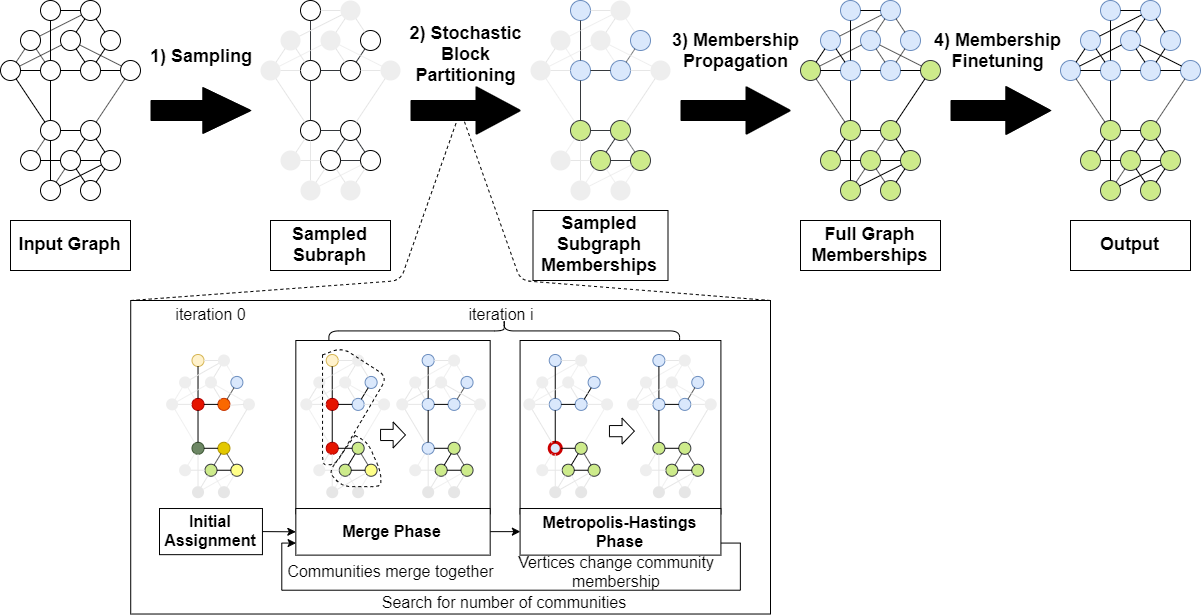}
  \caption{An illustration of the SamBaS community detection approach.}
  \label{fig:sbp-sampling}
\end{figure*}


In this section, we describe our four-step sampling-based stochastic block partitioning~(SamBaS) methodology and the sampling algorithms used in its evaluation.

\subsection{Overall Approach}\label{sec:approach}

We propose SamBaS, a four-step approach (illustrated in Figure~\ref{fig:sbp-sampling}) to accelerate community detection using sampling, as prototyped in~\cite{Wanye2019FastSampling}. Given a graph $G$ containing the set of vertices $V$ and edges $E$, a community detection algorithm run on $G$ produces a vertex-to-community assignment vector $A$. The goal of our approach is to perform community detection on a sampled graph $G^S$ and then extend the results to $G$, producing a vertex-to-community assignment $A^S$ such that it is of similar quality to $A$, in a much shorter amount of time. We adopt two measures of quality for $A$ - F1 Score for synthetic graphs where the true communities are known (see \S\ref{sec:synthetic_experiments}) and Partition Quality Score for real-world graphs where the true communities are not known (see \S\ref{sec:quality_metric}).

The four steps of our approach are outlined below:
\begin{enumerate}
    \item 
    A sampling algorithm is used to select $V^S$, a subset of vertices from the full graph. The sampled graph $G^S$ is then generated by adding only the edges that exist between vertices in $V^S$. Previous theoretical work has shown that the amount of community information provided by an edge is proportional to the degrees of the two vertices linked by said edge~\cite{Kao2019HybridInteractions}. Therefore, to increase the information content of the samples collected, we propose a thresholding scheme that enhances existing sampling algorithms. Under this scheme, vertices with a degree of less than 3 are excluded from the sample. In our experiments, this significantly improves the quality of results obtained (see \S~\ref{sec:thresholding}) by both raising the median F1 score and decreasing the variation in F1 score of the results obtained.
    \item Stochastic block partitioning (described in Figure~\ref{fig:sbp}) is applied to
    $G^S$, obtaining a community membership for every vertex in $V^S$. Because community detection algorithms tend to have superlinear computational complexities, we expect the time taken to perform community detection on $G^S$ to decrease by a factor larger than $\frac{|V|}{|V^S|}$.
    This step can be replaced with any other non-overlapping community detection algorithm, including a distributed implementation of SBP.
    \item The resulting community assignments for $V^S$ need to be propagated to the entire set of vertices $V$.
    First, the community membership of a vertex in $V$ is set to the community to which it has the most outgoing edges. Then, if the vertex is not directly connected to any vertices in $V^S$, a random community is assigned to it.
    \item 
    Because the above propagation step can lead to a sub-optimal 
    $A^S$,
    the community memberships of $V$ are further fine-tuned via a non-agglomerative vertex-level algorithm to produce the final vertex-to-community assignment $A^S$.
\end{enumerate}

SamBaS is heavily reliant on the quality of the sample on which community detection is run. A sample that does not preserve community structure will lead to poor vertex-to-community assignments for $V^S$, which will in turn lead to a low quality $A^S$. For this reason, the choice of sampling algorithm and sample size is very important.

\subsection{Sampling Algorithms}\label{sec:sampling_algs}

Here we describe the sampling algorithms that we incorporate into SamBaS.

\textbf{Expansion Snowball Sampling (ES):} The ES algorithm~\cite{Maiya2010SamplingStructure} is a non-probabilistic exploration-based sampling algorithm, based on the concept of the Expansion Factor~(EF). The algorithm keeps track of both the vertices in the current sample and the neighbors of the sample (vertices linked to the sample by outgoing edges). At each step, the vertex that leads to the highest increase in the sample's EF is selected, by selecting the vertex that contributes the most new neighbors to the sample. These calculations make ES sampling more computationally expensive than the remaining algorithms, but it has been shown to produce good results which can be extended to the full graph~\cite{Maiya2010SamplingStructure}.

\textbf{Maximum Degree Sampling (MD):} The MD algorithm is a natural extension of the thresholded sampling approach and aims to capture the most informative vertices. If vertices with high degrees contribute the most community-related information to the sample, then naturally, the best sampling strategy would be one that simply samples the vertices with the highest degrees. However, this method is likely to produce sampled graphs with a very different structure from the original graph, which intuitively should limit the quality of community detection results obtained with this algorithm. Good performance with this algorithm would suggest that vertex degree is the most important vertex property related to community structure.

The remaining three algorithms - Random Node Neighbor Sampling~(RNN), Uniform Random Sampling~(UR), and Forest Fire Sampling~(FF), are described in~\cite{Wanye2019FastSampling}.

\section{Experimental Setup}

In this section, we describe the graphs used for our experimental study as well as the hardware and software used to run these experiments.

\subsection{Real-World Graphs}\label{sec:real_graphs}

We select a set of 10 web graphs, where vertices represent webpages and edges represent links between them, from the Network Data Repository~\cite{Rossi2015TheVisualization}.
Like most real-world scenarios, the true community memberships for vertices in these graphs is unknown.
We describe our solution to evaluating community detection results on these graphs in~\S\ref{sec:quality_metric}.

Table~\ref{tab:real_graphs} presents the selected graphs. With respect to the known published SBP results in~\cite{Kao2017StreamingPartition,Ghosh2018ScalableVite,Uppal2017ScalablePartition}, these are the largest graphs in terms of both $|V|$ and $|E|$.

\begin{table}[ht!]
\centering
\caption{Selected real-world graphs.}
\label{tab:real_graphs}
\begin{tabular}{cccrr}
\hline
ID & Name & Directed? & $|V|$ & $|E|$ \\
\hline
R1 & web-NotreDame & True & 326k & 1497k \\
R2 & web-arabic-2005 & False & 164k & 1747k \\
R3 & web-Stanford & True & 282k & 2312k \\
R4 & web-italycnr-2000 & True & 326k & 3216k \\
R5 & web-baidu-baike-related & True & 416k & 3284k \\
R6 & web-wikipedia2009 & False & 1864k & 4507k \\
R7 & web-google-dir & True & 876k & 5105k \\
R8 & web-it-2004 & False & 509k & 7178k \\
R9 & web-BerkStan-dir & True & 685k & 7601k \\
R10 & web-uk-2005 & False & 130k & 11744k \\
\hline
\end{tabular}
\end{table}





\subsection{Synthetic Graphs}\label{sec:synthetic_graphs}\label{sec:graph_generation}

To study how individual graph parameters affect SamBaS, we need the ability to individually alter these parameters with minimal side effects on the graph structure. Additionally, we need ground truth labels to be available in order to measure community detection accuracy easily.
%
%
%
To satisfy these requirements, we generate directed graphs based on the DSCBM~\cite{Karrer2011StochasticNetworks} using the \verb|graph-tool| software library~\cite{Peixoto2014TheLibrary}, similar to those generated for the IEEE/Amazon/MIT Streaming Graph Challenge~\cite{Kao2017StreamingPartition}. This ensures that the vertex community memberships are known and that the graph has a well-defined community structure.

In generating our graphs, we vary the following six parameters $P$ one at a time, leading to six sets of synethetic graphs as shown in Table~\ref{tab:synthetic_graphs}:

\begin{enumerate}
    \item Number of vertices $|V|$
    \item Number of communities $C$
    \item Standard deviation of community sizes $\sigma$
    \item Strength of community structure $s$, measured by the ratio of the number of edges within communities to the number of edges between communities.
    \item Strength of high-degree vertices $d^{max}$
    \item Density $\rho$
\end{enumerate}

Note that due to the probabilistic nature of the graph generator, many of these parameters are controlled indirectly, leading to some difference between the expected values of these parameters and the final values in the generated graph. In addition, due to the interactions between inputs during graph generation, we do sometimes see significant changes in an unintended parameter (most notably in the graphs where the number of communities is varied), but to the best of our knowledge completely eradicating such interactions is impossible without breaking the parameters of the underlying DCSBM.


\begin{table}[h!]
\centering
\caption{Parameters of Synthetically Generated Graphs}
\label{tab:synthetic_graphs}
\resizebox{\columnwidth}{!}{%
\begin{tabular}{ccrrrrrrr}
\hline
ID & $P$ & \multicolumn{1}{c}{$|V|$} & \multicolumn{1}{c}{$|E|$} & \multicolumn{1}{c}{$C$} & \multicolumn{1}{c}{$\sigma$} & \multicolumn{1}{c}{$s$} & \multicolumn{1}{c}{$d^{max}$} & \multicolumn{1}{c}{$\rho$} \\
\hline
S1  & $|V|$     & 64k   & 0.3M   & 268  & 150  & 3.85 & 3.6k  & 6.6E-05 \\ 
S2  & $|V|$     & 130k  & 0.6M   & 380  & 212  & 3.90  & 12.9k & 3.6E-05 \\ 
S3  & $|V|$     & 260k  & 1.3M  & 537  & 288 & 3.84 & 9.6k  & 1.9E-05 \\ 
S4  & $|V|$     & 522k  & 2.7M  & 760  & 411 & 3.85 & 31.2k & 1.0E-05 \\ 
S5  & $|V|$     & 1039k & 5.4M  & 1075 & 588 & 3.87 & 21.2k & 5.0E-06 \\ 
S6  & $|V|$     & 2084k & 11.2M & 1521 & 834 & 3.87  & 63.3k & 2.6E-06 \\ 
\hline
S7  & $C$ & 301k  & 1.5M  & 128  & 1493 & 3.91 & 25.1k & 1.6E-05 \\ 
S8  & $C$ & 303k  & 1.5M  & 256  & 682 & 3.75  & 18.7k & 1.6E-05 \\ 
S9  & $C$ & 303k  & 1.5M  & 512  & 368  & 3.89 & 16.9k & 1.6E-05 \\ 
S10 & $C$ & 304k  & 1.5M  & 1024 & 189 & 3.94 & 9.5k  & 1.6E-05 \\ 
S11 & $C$ & 308k  & 1.5M  & 2048 & 93  & 3.92 & 6.4k  & 1.6E-05 \\ 
\hline
S12 & $\sigma$   & 308k  & 1.5M  & 582  & 186 & 3.27 & 10.9k & 1.6E-05 \\ 
S13 & $\sigma$   & 305k  & 1.5M  & 582  & 273 & 3.63 & 9.5k  & 1.6E-05 \\ 
S14 & $\sigma$   & 304k  & 1.5M  & 582  & 318 & 3.85 & 11.4k & 1.7E-05 \\ 
S15 & $\sigma$   & 302k  & 1.5M  & 582  & 362 & 4.15 & 14.1k & 1.6E-05 \\ 
S16 & $\sigma$   & 300k  & 1.6M  & 582  & 414 & 4.49 & 17.5k & 1.7E-05 \\ 
\hline
S17 & $s$     & 307k  & 1.4M  & 582  & 327 & 1.32 & 10.9k & 1.5E-05 \\ 
S18 & $s$     & 303k  & 1.4M  & 582  & 341 & 2.67 & 10.9k & 1.6E-05 \\ 
S19 & $s$     & 303k  & 1.4M  & 582  & 325 & 3.89 & 9.1k  & 1.5E-05 \\ 
S20 & $s$     & 303k  & 1.5M  & 582  & 318 & 5.13 & 12.8k & 1.6E-05   \\
S21 & $s$     & 302k  & 1.5M  & 582  & 336 & 6.58 & 13.6k & 1.6E-05 \\ 
\hline
S22 & $d^{max}$   & 301k  & 1.3M  & 582  & 326 & 3.90 & 4.8k  & 1.4E-05 \\ 
S23 & $d^{max}$   & 301k  & 1.3M  & 582  & 320 & 3.86 & 5.6k  & 1.5E-05 \\ 
S24 & $d^{max}$   & 303k  & 1.4M  & 582  & 319 & 3.87 & 7.6k  & 1.6E-05 \\ 
S25 & $d^{max}$   & 306k  & 1.6M  & 582  & 314 & 3.85 & 14.7k & 1.7E-05 \\ 
S26 & $d^{max}$   & 305k  & 1.7M  & 582  & 348 & 4.03  & 71.5k & 1.8E-05 \\ 
\hline
S27 & $\rho$     & 110k  & 55.2M & 336  & 183 & 3.88  & 20.7k & 4.6E-03 \\ 
S28 & $\rho$     & 108k  & 41.3M & 336  & 185 & 3.91 & 15.7k & 3.5E-03 \\ 
S29 & $\rho$     & 105k  & 27.9M & 336  & 182 & 3.91 & 9.8k  & 2.5E-03 \\ 
S30 & $\rho$     & 99k   & 14.3M & 336  & 175 & 3.92 & 5.4k  & 1.5E-03 \\ 
S31 & $\rho$     & 56k   & 0.3M   & 336  & 113 & 4.03   & 0.1k   & 1.1E-04 \\ 
\hline
\end{tabular}
}
\end{table}

\subsubsection{Similarity to Real-World Graphs}

\begin{figure}[h]
  \centering
  \includegraphics[width=\columnwidth]{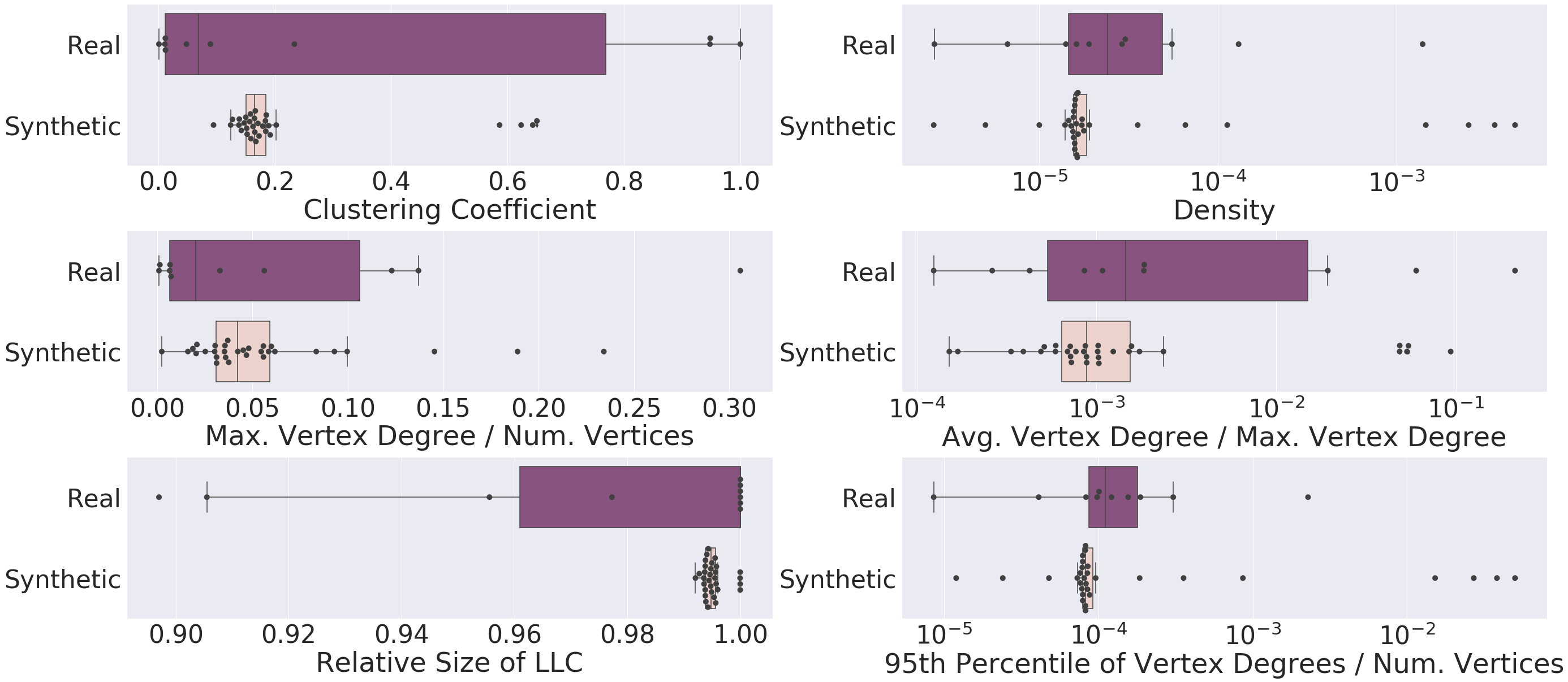}
  \caption{Comparison of distributions of parameters of real-world graphs and our synthetically generated graphs. The synthetic graph parameters are clustered around the average parameter values for the real-world graphs, while still having enough of a spread to sweep over the range of the parameter values for the real-world graphs. Note the log scale on the x-axis for plots on the right.}
  \label{fig:graph_comparison}
\end{figure}

To ensure that the results obtained from the 
synthetically generated graphs 
are applicable to real-world data, we compare our graphs to the selected web graphs. 
However, because we do not know the real community structure of the web graphs, we cannot compare parameters like $C$, $\sigma$, or $s$ between the two sets of graphs. Additionally, many graph features depend on and scale with other features. As an example, the number of edges in a graph and the maximum vertex degree tend to increase as the number of vertices in the graph increases. Therefore, in order to compare our synthetic graphs to these web graphs, we heavily utilize ratios of parameters. As can be seen in Figure~\ref{fig:graph_comparison}, our synthetic graphs are similar to the chosen web graphs with respect to the following parameters:

\begin{itemize}
    \item Clustering coefficient
    \item Density ($\rho$) of the graph
    \item Ratio of maximum vertex degree $d^{max}$ to number of vertices $|V|$
    \item Ratio of average vertex degree $d^{avg}$ to maximum vertex degree $d^{max}$
    \item Relative size of largest connected component (as defined by $\frac{|V^C|}{|V|}$, where $|V^C|$ is the number of vertices in the largest connected component~(LLC)
    \item Ratio of 95th percentile of vertex degrees $d^{95}$ to number of vertices $|V|$
\end{itemize}

For each parameter, the values of the synthetic graphs are clustered around the average value for the real-world graphs, showing that our graphs are similar to the average web graph. The spread of the parameter values for the real-world graphs in Figure~\ref{fig:graph_comparison} 
shows that we have enough variation to sweep over the range of values that each parameter takes in real-world graphs.

\subsection{Hardware and Software Infrastructure}\label{sec:hardware}

We performed all of our experimental runs on the
Intel\textregistered DevCloud, using computational nodes with 192~GB of memory and two Intel\textregistered Xeon\textregistered Gold 6128 processors, each supporting 12 concurrent threads. We make use of graph-tool's built-in OpenMP-based parallelization of SBP and implement the sampling approach code in serial Python using the NumPy library~\cite{Oliphant2006GuideNumPy,vanderWalt2011TheComputation}.
\label{sec:experimental_setup}

\section{Experiments on Synthetic Graphs}\label{sec:synthetic_experiments}

In this section, we describe a set of experiments performed on the synthetic graphs in Table~\ref{tab:synthetic_graphs}. For every graph, we run SamBaS using each of the five sampling algorithms from \S\ref{sec:sampling_algs} at the 50\%, 30\% and 10\% sample sizes. To provide a baseline for comparison, we 
run the SBP algorithm on the full graph without sampling. Additionally, we run SamBaS \emph{without} thresholding on graphs S1 to S6 using the ES, RNN, UR and FF sampling algorithms. Both the baseline and the SBP step of our sampling approach are implemented via the same
call to the graph-tool~\cite{Peixoto2014TheLibrary} library.

Due to the stochastic nature of SBP, 
the algorithm can get stuck in a local minimum, leading to sub-optimal results on some runs and a large variance in the quality of community detection results. To reduce this variance, it is common practice to execute the algorithm multiple times independently, and select the best result~\cite{Gelman2013BayesianAnalysis}. We incorporate this technique into our approach by performing 5 runs for each experiment, where each run consists of 2 independent executions of SBP.

We evaluate two aspects of SamBaS: the speedup achieved and the quality of results. We evaluate the speedup by comparing the execution times of the runs with SamBaS to the baseline runs. Similar to the Graph Challenge~\cite{Kao2017StreamingPartition}, we evaluate the quality of results by matching the communities obtained during the selected run to the true community memberships for each vertex using the Hungarian algorithm, and then calculating the F1 score, where $\texttt{F1 Score} = \frac{2 \times \texttt{Precision} \times \texttt{Recall}}{\texttt{Precision} + \texttt{Recall}}$.

\subsection{Results for Thresholded Sampling}~\label{sec:thresholding}

\begin{figure}[h!]
  \centering
  \includegraphics[width=\columnwidth]{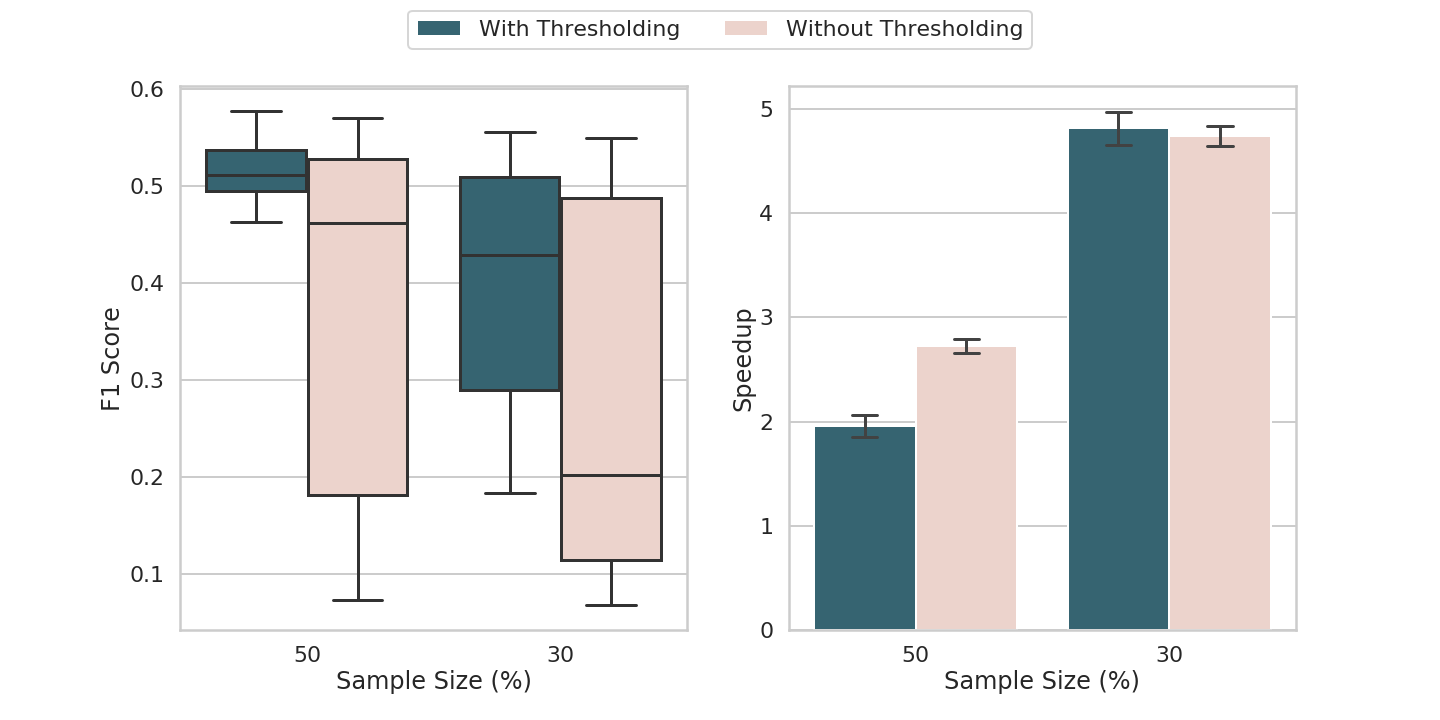}
  \caption{Comparison of F1 scores (left) and average speedups compared to performing community detection without sampling (right) obtained with and without thresholding at the 50\% and 30\% sample sizes on graphs S1-S6. Results combine data from the uniform random, random node neighbor, forest fire, and expansion snowball sampling algorithms.}
  \label{fig:thresholding}
\end{figure}

We evaluate thresholded SamBaS on graphs S1-S6 and compare it to SamBaS without thresholding. The MD sampling algorithm is excluded from these experiments since thresholding has no effect on it. 
As seen in Figure~\ref{fig:thresholding}, thresholding leads to higher average F1 scores and lower variation in F1 scores at both the 50\% and 30\% sample sizes, when compared to SamBaS without thresholding.
However, it also leads to a drop in average speedup from 2.7$\times$ to 2.0$\times$ at the 50\% sample size (though at the 30\% sample size, average speedup remains roughly unchanged).

At the 10\% sample size, the difference between runs with and without thresholding is not significant because a 10\% sample size is generally not large enough to capture all of the necessary structural information from the graph.

\subsection{Results for Individual Experiments}

We identify trends in the comparative accuracy and speedups gained using thresholded SamBaS for each experiment. As seen in 
Appendix~A, MD sampling generally outperformed the other sampling algorithms. As such, for the sake of brevity, we only include results using the MD sampling algorithm for the remainder of this section.


\subsubsection{Trends in Result Quality}

\begin{figure}[ht!]
  \centering
  \includegraphics[width=\columnwidth]{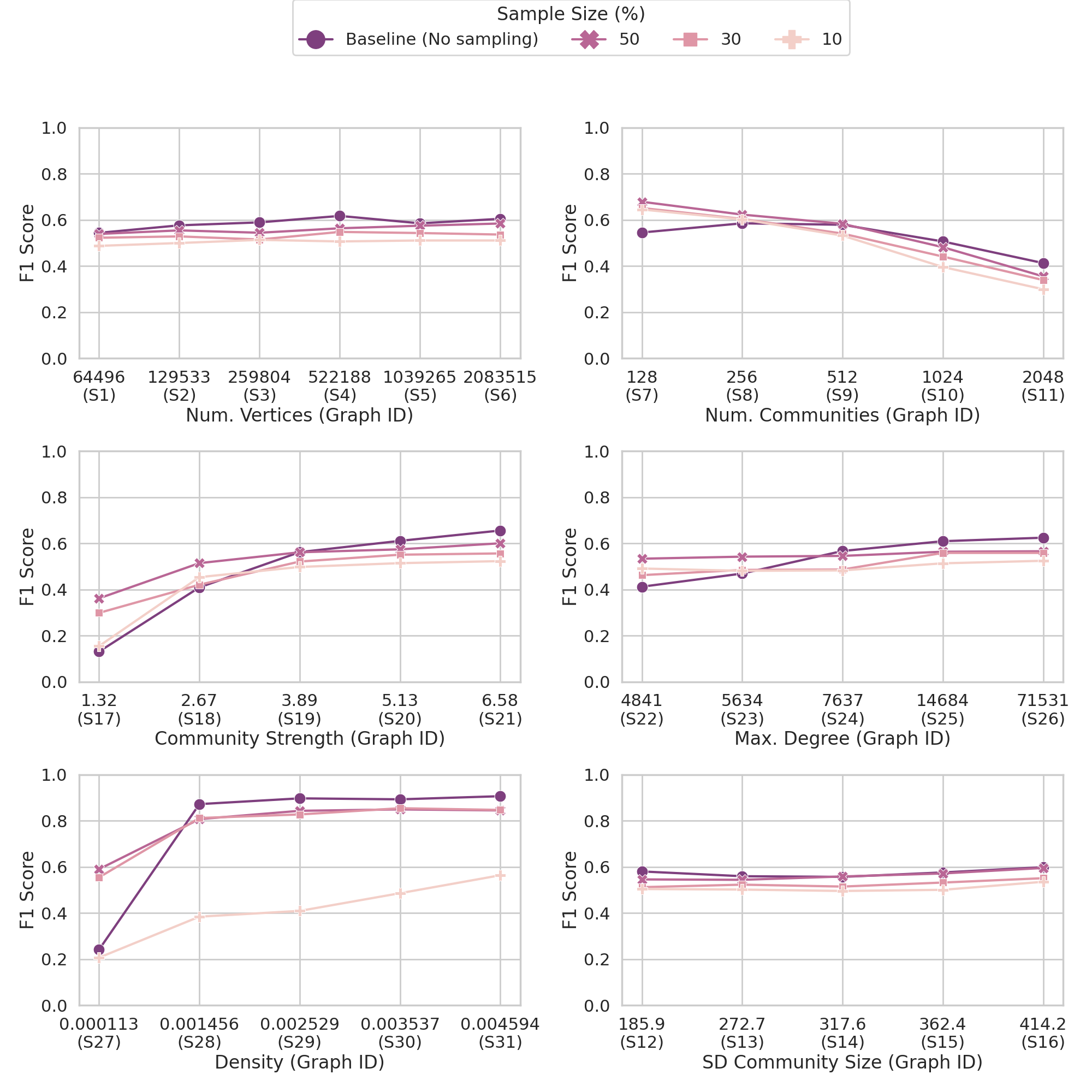}
  \caption{Comparison of F1 scores obtained over the baseline (stochastic block partitioning without sampling) for all 6 experiments using maximum degree sampling.}
  \label{fig:synthetic_f1}
\end{figure}

First, we identify trends with respect to community detection accuracy.
Though we only show the trends obtained with MD sampling in Figure~\ref{fig:synthetic_f1}, we find that they are closely replicated by the other sampling algorithms. The results discussed here are relative to the baseline SBP (i.e., 100\% sample size). When SamBaS achieves ``good'' F1 scores, the F1 scores achieved with sampling are comparable to those achieved by the baseline SBP. When SamBaS leads to ``poor'' F1 scores, they are considerably lower than those achieved by the baseline.

Empirically, the scale of the graph (i.e., number of vertices) does not affect the quality of results obtained via sampling.
Although SamBaS leads to relatively poor F1 scores on graphs S3 (260k vertices) and S4 (522k vertices), it does achieve good results on the smaller graphs S1 and S2 and on the larger graphs S5 and S6. 
Overall, the F1 scores achieved with and without sampling increase with the size of the graph.
One factor that contributes to this is the sublinear increase in the number of communities with respect to $|V|$, 
where F1 scores are higher in graphs with fewer communities.


In most of the remaining experiments, including those that varied the number of communities $C$ (graphs S7-S11),
the strength of the communities $s$ (graphs S17-S21),
the maximum vertex degree $d^{max}$ (S22-S26),
and density $\rho$ (graphs S27-S31),
our approach works best when the aforementioned parameters are low. So much so that, on graphs S7-S9 ($C \leq 512$), S17-S18 ($s \leq 2.67$), S22-S23 ($d^{max} \leq 5634$), and S27 ($\rho = 0.0001$), SamBaS achieves \textit{higher} F1 scores than the baseline. In S17 and S27, F1 scores with SamBaS at the 50\% sample size are over twice as high as F1 scores without sampling.

On the other hand, experiments with community size variation $\sigma$ (S12-S16),
show the reverse trend, and SamBaS is unable to achieve higher F1 scores than the baseline, with the possible exception of graph S14, where the average F1 score with sampling is very slightly higher at the 50\% sample size.

\subsubsection{Trends in Speedup}

\begin{figure}[ht!]
  \centering
  \includegraphics[width=\columnwidth]{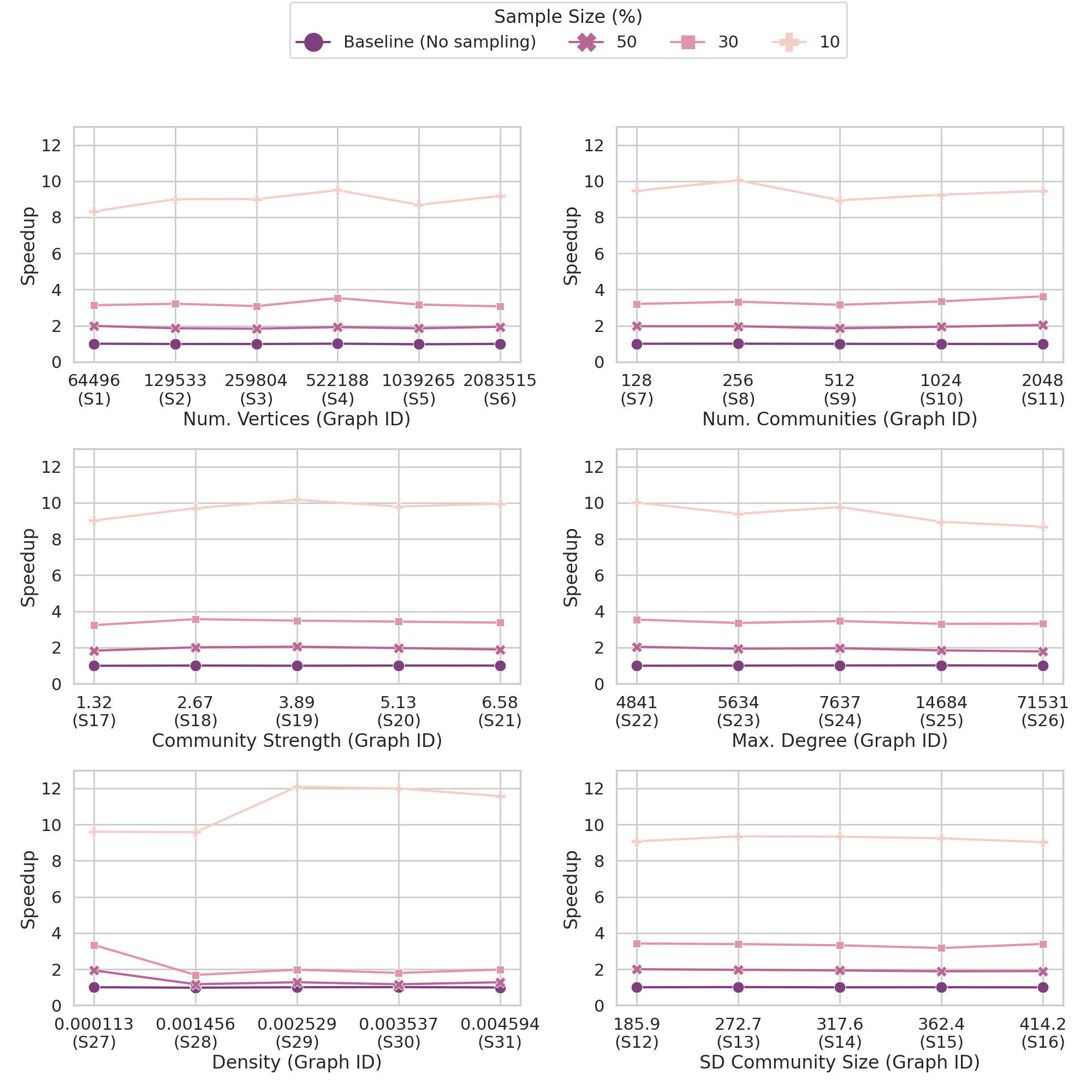}
  \caption{Comparison of speedup obtained over the baseline (stochastic block partitioning without sampling) for all 6 experiments using maximum degree sampling.}
  \label{fig:synthetic_speedup}
\end{figure}

Next, we identify the trends in speedup with respect to the parameters modified in each experiment. These trends are summarized in Figure~\ref{fig:synthetic_speedup}.

For the majority of the studied graph parameters, including graph size $|V|$ (graphs S1-S6),
number of communities $C$ (graphs S7-S11),
the standard deviation in community size $\sigma$ (graphs S12-S16),
and the strength of community structure $s$ (graphs S17-S21),
our results indicate that varying the parameter does not affect speedup by a significant amount. The 10\% sample size is the exception, where speedup generally increases as $|V|$ and $s$ increase.  

On the other hand, maximum vertex degree $d^{max}$ (graphs S22-26),
and density $\rho$ (graphs S27-S31),
do affect speedup. As $d^{max}$ increases, the percentage of edges present in the sampled graph increases, causing the SBP step of our sampling approach to take longer, and the average speedup obtained to decrease. 

Experiments with $\rho$ show a more complicated pattern; speedup at the 10\% sample size increases as density increases, but at the 50\% and 30\% sample sizes, speedup is highest at the lowest $\rho$ (graph S27), after which it drops drastically and then gradually increases as $\rho$ increases. These results can be explained by a combination of two factors. First, SBP is faster at lower densities because the algorithm has to process fewer edges. This means that at lower densities, and especially when the low density is combined with a low sample size, the amount of time taken to propagate the community memberships from the sampled graph to the remaining vertices, and to finetune the results, becomes more significant. The second factor is that, similarly to the $d^{max}$ experiments, the percentage of edges present in the sampled graph increases as the graph density increases, leading to lower speedups at higher sample sizes.

\subsection{Trade-offs between Result Quality and Speedup}

\begin{figure}[ht!]
  \centering
  \includegraphics[width=\columnwidth]{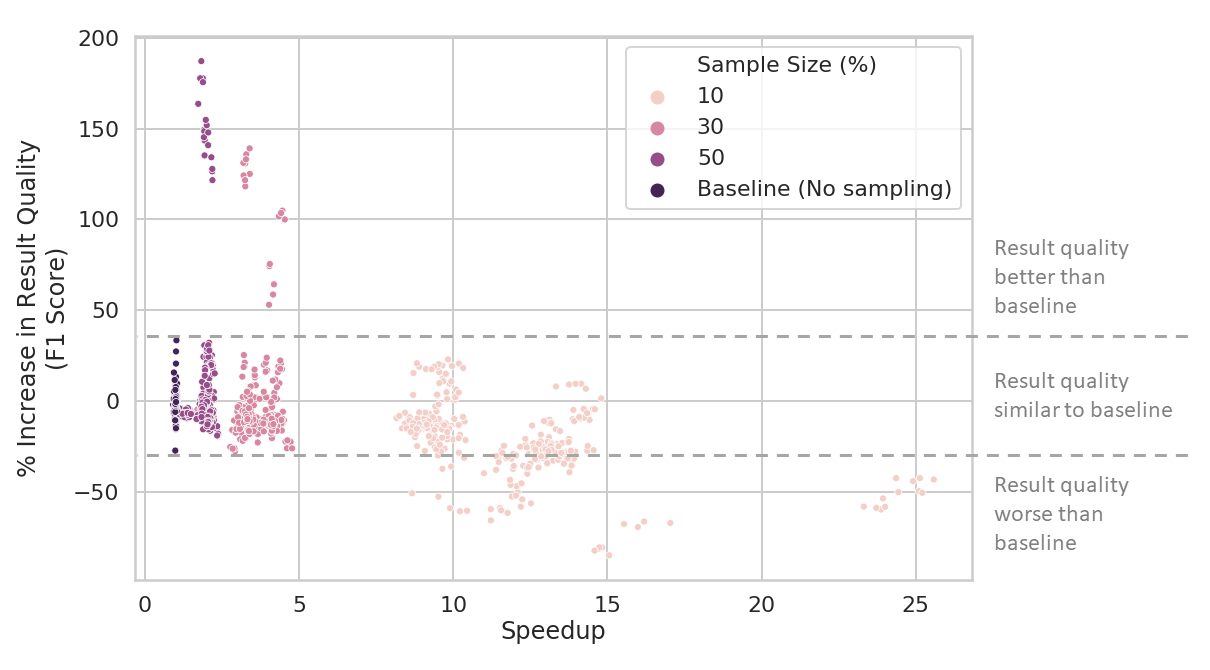}
  \caption{Trade-off between result quality and speedup with our sampling approach on a set of realistic synthetic graphs (see \S\protect\ref{sec:synthetic_graphs}) using the best-performing sampling algorithms (see 
  Appendix~A.1). Each point represents one run with a particular combination of sample size, sampling algorithm, and graph. Metrics are calculated against the average baseline result on a per-graph basis. The results from the baseline algorithm runs demonstrate the variation in result quality that is inherent to stochastic block partitioning. We achieve a speedup up to 10$\times$ without significantly affecting the result quality. Alternatively, we achieve greater than 150\% improvement in result quality on certain graphs at a more modest speedup.}
  \label{fig:intro}
\end{figure}

Finally, we examine the results from all 31 synthetic graphs and identify the trade-offs incurred between result quality and speedup using SamBaS. As seen in Fig.~\ref{fig:intro}, the majority of results obtained at the 50\% and 30\% sample sizes are comparable to the results obtained without sampling. All the cases where SamBaS performs better than the baseline in terms of result quality occur at these two sample sizes. However, they result in relatively lower speedups of up to 5X. On the other hand, decreasing the sample size to 10\% does lead to a reduction in result quality in some graphs. In many graphs, however, SamBaS at the 10\% sample size obtains speedups of up to 10X without significantly affecting result quality, making smaller sample sizes viable in situations where accuracy is not of the utmost importance.

\subsection{Discussion}

One would intuitively expect that an approach based on data reduction would work best for simpler graphs, where the community structure is more obvious. For our experiments, this would translate to better F1 scores in comparison to the baseline when the baseline F1 scores are high. However, for many experiments, including $s$, $d^{max}$, and $\rho$, SamBaS performs much better when the graph's communities are harder to accurately detect. Additionally, although $\sigma$ does not have a strong effect on baseline F1 scores, our sampling approach performs better comparatively when $\sigma$ is high, despite the fact that the community sizes are more uneven, making it harder to perform community detection on those graphs.

We postulate that the explanation for this phenomenon lies in the relative amount of information captured via sampling. Intuitively, graphs that present a harder community detection problem would contain fewer vertices that provide useful structural information. For example, when the graph is too sparse, several vertices could have unclear community memberships because they're connected to only 1 other vertex. Similarly, when the community structure is not strong, vertices may be equally strongly connected to several communities. In such cases, a good sampling algorithm could narrow the search space for the MCMC process down to just the vertices that provide useful structural information, leading to good community detection results.

Additionally, narrowing down the search space could make it easier for the SBP algorithm to find the globally optimal solution, provided that the search space still contains enough structural information. This could explain how SamBaS sometimes leads to better community detection results than the baseline; the baseline algorithm simply gets stuck in a locally optimal solution.


Several avenues for future work present themselves based on these results. One would be a factorial experiment design to identify more complex performance patterns that exist due to interactions between graph parameters. Another is developing parallel or even distributed versions of the sampling algorithms, which could lead to further speedup gains, especially on large graphs. Lastly, a deep dive into our approach on the graphs where SamBaS improved result quality could lead to a more concrete explanation of the aforementioned phenomenon, and maybe even provide a theoretical intuition into how to achieve the same effect on other graphs.


\section{Experiments on Real-World Graphs}

In this section, we describe the experiments performed on the set of real-world web graphs from Table~\ref{tab:real_graphs}.

\subsection{Evaluating Community Detection Without Known Ground Truth}\label{sec:quality_metric}


In real-world scenarios, graphs with known true community memberships are incredibly scarce. Even when they are available, as in the Stanford Large Network Dataset Collection~\cite{Leskovec2014SNAPCollection}, the memberships are often overlapping. This makes it difficult to evaluate the quality of non-overlapping community detection results such as those obtained via SBP, since straightfoward metrics such as accuracy and F1 score cannot be reliably calculated.


\begin{table}[ht!]
\centering
\caption{Evaluation of Community Detection in Graphs Without Ground Truth.}
\label{tab:related_work_evaluation}
\resizebox{\columnwidth}{!}{%
\begin{threeparttable}
\begin{tabular}{c|c|c|c}
\hline
\multirow{2}{*}{Paper} & \multirow{2}{*}{Algorithm(s)} & \multicolumn{2}{c}{Evaluation}\\
\cline{3-4}
& & Metric & F1 Correlation\\
\hline
Lu~\cite{Lu2015ParallelDetection} & Louvain & Modularity &  \\
Que~\cite{Que2015ScalableAlgorithm} & Louvain & Modularity &  \\
Halappanavar~\cite{Halappanavar2017ScalableGrappolo} & Louvain & Modularity &  \\
Naim~\cite{Naim2017CommunityGPU} & Louvain & Modularity &  \\
Uppal~\cite{Uppal2017ScalablePartition} & SBP & N/A &  \\
Ghosh~\cite{Ghosh2018ScalableVite,Ghosh2018DistributedDetection} & Louvain & Modularity &  \\
Ghosh~\cite{Ghosh2019ScalingClustering} & FTR\tnote{1} & Modularity &  \\
Maiya~\cite{Maiya2010SamplingStructure} & GN, NLE, CNM\tnote{2} & N/A & \\
Stanley~\cite{Stanley2018CompressingNodes} & Louvain, SBP & N/A\tnote{3} &  \\
\hline
This Manuscript & SBP & MDL-based & \checkmark \\
\hline
\end{tabular}
\begin{tablenotes}
\item[1] Fast-Tracking Resistance (FTR).
\item[2] Girvan-Newman (GN), Newman's Leading Eigenvector (NLE), Clauset-Newman-Moore (CNM).
\item[3] Results on graphs without known ground truth were evaluated by calculating NMI against the baseline algorithm.
\end{tablenotes}
\end{threeparttable}}
\end{table}

As seen in Table~\ref{tab:related_work_evaluation}, the majority of community detection literature either does not attempt to evaluate community detection results on graphs with no known ground truth, or employs the modularity metric.
Modularity measures how strongly vertices within communities are connected, as compared to a random graph, making it an intuitive choice for evaluating community detection. However, it has been shown that high modularity values do \emph{not} necessarily correspond to the most semantically accurate community assignments~\cite{Fortunato2010CommunityGraphs}. 
In~\cite{Bianconi2009AssessingStructure}, it was further shown that modularity can fail to differentiate between graphs with very different degrees of separation between communities. 

To overcome these weaknesses, we devise an alternative evaluation metric based on the minimum description length $H$ of the DCSBM. By basing our evaluation on $H$, our metric takes into account both the number of communities obtained and the difference in degree distribution between communities via log-likelihood calculations (see \S\ref{sec:sbp}). Furthermore, the goal of SBP is to minimize $H$. As such, comparing the values of $H$ obtained with and without sampling is a natural way to evaluate distinct community detection approaches that use the SBP algorithm.


Our metric, the \textit{Partition Quality Score}~($PQS$), is defined as the amount of the blockmodel minimum description length that has been compressed through performing community detection. $PQS$ is given by the following equation:

\begin{equation}\label{eq:PQS}
    PQS = \frac{H^{max} - H}{H^{max}},
\end{equation}

\noindent
where $H^{max}$ is the base value of $H$ for a given graph (when every vertex is considered to be a separate community), and $H$ is the minimum description length obtained via running a community detection algorithm~(SBP or SamBaS) on the same graph.

\begin{figure}[ht!]
  \centering
  \includegraphics[width=0.9\columnwidth]{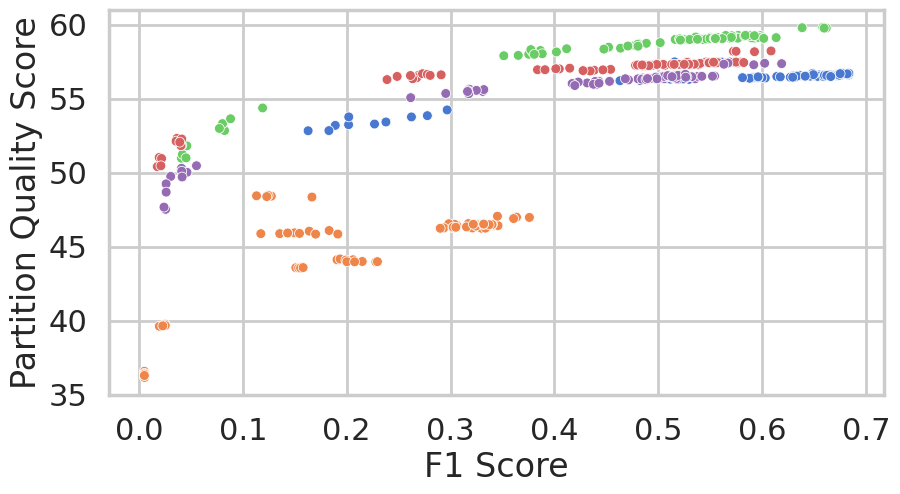}
  \caption{The relationship between community detection accuracy based on truth (F1 Score) and Partition Quality Score, with points from community detection run on randomly selected synthetic graphs (indicated by color) across different sampling approaches and sample sizes. Partition Quality Score increases as community detection accuracy based on truth (F1 Score) increases, but at different trajectories for each graph.}
  \label{fig:metrics_different_graphs}
\end{figure}

To evaluate $PQS$ and modularity, we collect both metrics for our synthetic experimental runs, as articulated in \S\ref{sec:synthetic_experiments}, and compare them to the obtained F1 scores. As Figure~\ref{fig:metrics} shows, modularity does \emph{not} correspond well to the F1 scores obtained. In fact, for roughly half of our data points, modularity has an inverse relationship with F1 score. This is consistent with the findings in~\cite{Yang2015DefiningGround-truth}, which show that modularity has an inverse relationship with several measures of community quality. On the other hand, $PQS$ has a much more direct relationship with F1 score. However, as seen in Figure~\ref{fig:metrics_different_graphs}, the trajectory at which $PQS$ changes with F1 score is different for each graph. Thus $PQS$ can be used to compare accuracy on the same graph, but not across graphs.

\subsection{Experiments}

To evaluate SamBaS on real-world graphs, we adopt a similar approach to the one described in \S\ref{sec:synthetic_experiments} but with fewer parameters. For every graph in Table~\ref{tab:real_graphs}, we run our sampling approach using the three best-performing sampling algorithms from \S\ref{sec:synthetic_experiments} --- MD, RNN and ES --- but with thresholding. 
We only run these algorithms at the 30\% sample size as a compromise between speedup and quality of community detection. To provide a baseline for comparison, we also run the SBP algorithm on each full graph without sampling. 

We evaluate two aspects of SamBaS; the speedup gained and the quality of results. We evaluate the speedup by comparing the execution times of the SamBaS runs to the baseline runs and the quality of results using the \emph{Partition Quality Score} metric described above. For the results on all experiments performed, refer to 
Appendix~B.

\subsection{Results on Web Graphs}

\begin{figure}[h!]
  \centering
  \includegraphics[width=0.45\textwidth]{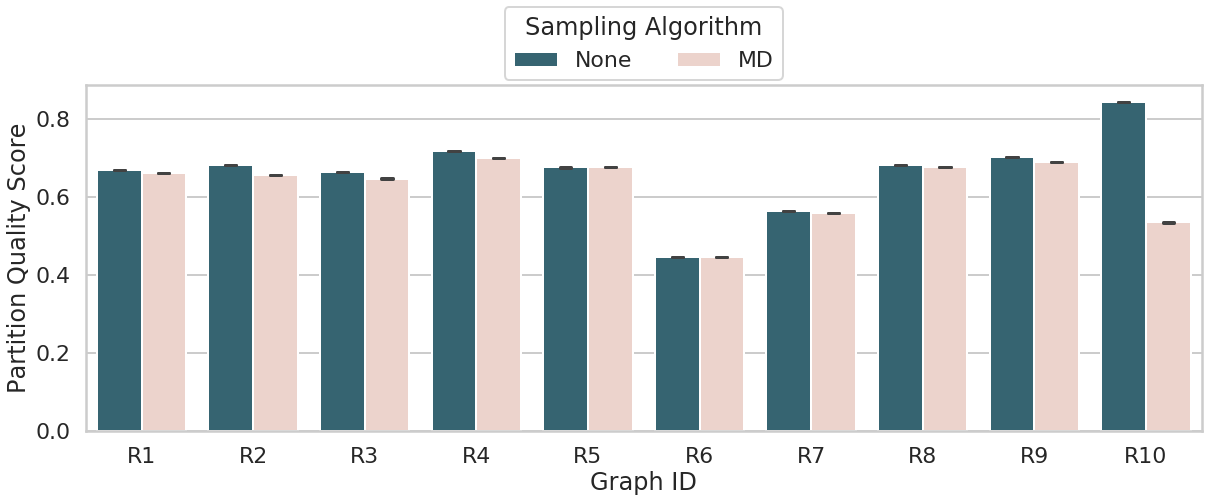}
  \caption{Comparison of the Partition Quality Score achieved on 10 web graphs with and without our sampling approach (higher is better). The maximum degree sampling algorithm (MD) was ran at a 30\% sample size.}
  \label{fig:real_cl}
\end{figure}

Similar to the results on synthetic graphs, we find that MD sampling generally produces the best results on real-world graphs. As Figure~\ref{fig:real_cl} shows, the quality of results using SamBaS is very similar to the baseline. In fact, for all 10 real-world graphs, our sampling approach achieves a $PQS$ within just 0.05 of the baseline.

\begin{figure}[h!]
  \centering
  \includegraphics[width=\columnwidth]{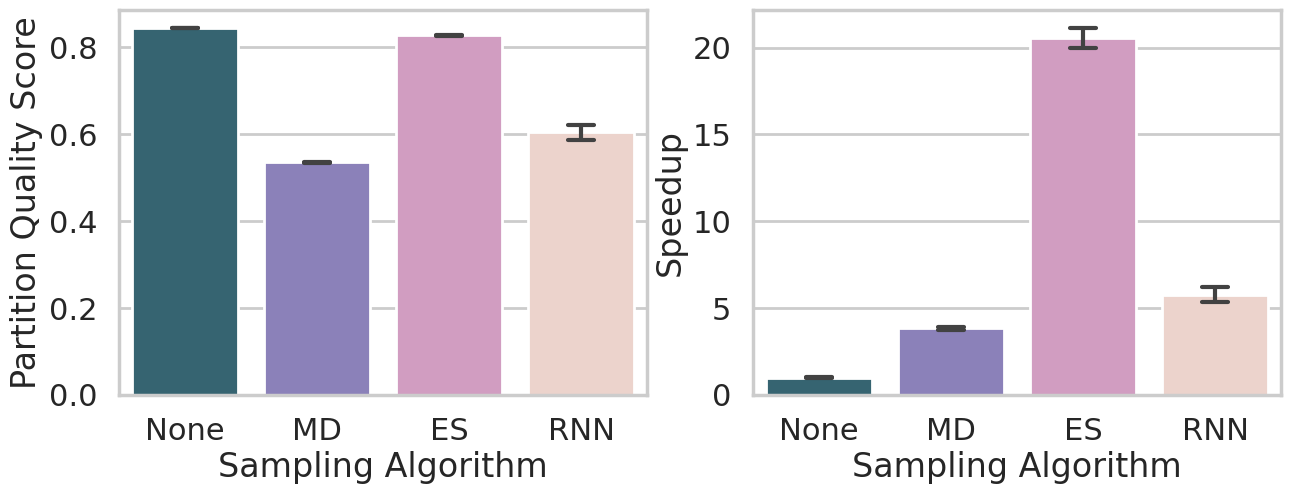}
  \caption{Comparison of the Partition Quality Score (left, higher is better) and average speedups (right) achieved on graph R10 (web-uk-2005) using the baseline, maximum degree (MD), expansion snowball (ES), and random node neighbor (RNN) sampling algorithms at the 30\% sample size.}
  \label{fig:real_r10}
\end{figure}

The most intriguing result occurs on graph R10. It is the only graph on which there is a large difference in $PQS$ between the tested sampling algorithms; it is also the only graph on which MD sampling leads to a poor $PQS$. However, as shown in Figure~\ref{fig:real_r10}, ES sampling achieves good $PQS$ results on the same graph. This outlier result can be explained by the composition of the communities within the R10 graph. Because MD sampling only samples the vertices with the highest degree, communities comprised of only low-degree vertices will \emph{not} be present in the sample, and thus, cannot be identified using our sampling approach. Upon examination of the communities obtained without sampling on R10, we find that only 37.5\% of them contain vertices with a high-enough degree to be selected by the MD algorithm at the 30\% sample size. The ES sampling algorithm, which prioritizes exploration of the graph, is unaffected by this phenomenon. The RNN algorithm, while performing better than MD, is also prone to undersampling such communities when the number of communities with high-degree vertices is large.

\begin{figure}[h!]
  \centering
  \includegraphics[width=0.45\textwidth]{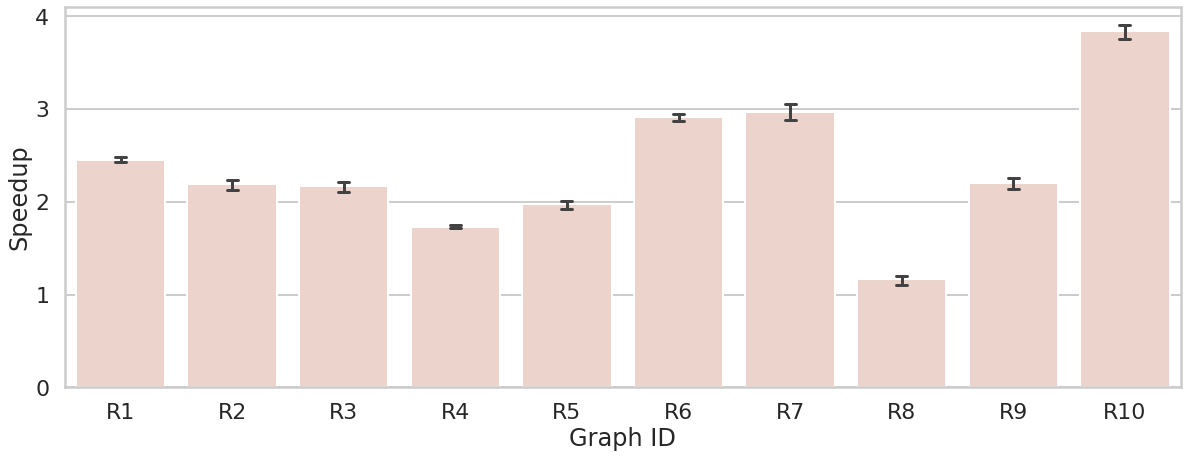}
  \caption{Comparison of the speedups obtained on 10 web graphs with our sampling approach over the baseline, using the maximum degree (MD) sampling algorithm at the 30\% sample size.}
  \label{fig:real_speedup}
\end{figure}

With MD sampling at the 30\% sample size, SamBaS achieves an average speedup of 2.36$\times$ across the web graphs R1-R10. As Figure~\ref{fig:real_speedup} shows, there is a relatively wide range of obtained speedups from only 1.16$\times$ on R8 to 3.84$\times$ on the much denser R10. RNN and ES sampling regularly result in higher speedups, most notably with ES achieving a speedup of over 20$\times$ on R10 at the same sample size. This is largely due to ES sampling producing much sparser sampled graphs, which results in more rapid blockmodel inference.

\subsection{Discussion}

The outlier graph R10 shows that the choice of sampling algorithm depends on the structure of the graph. While our synthetic graphs do not capture the behavior of R10, our own experiments on the Graph Challenge~\cite{Kao2017StreamingPartition} graphs showed a similar pattern to R10, where ES sampling produces better quality results than the other sampling algorithms, including RNN. This raises the question of whether or not other graph parameters exist that similarly affect the choice of sampling algorithm.

On the whole, however, the results in this section provide strong evidence that SamBaS works well on real-world graphs. This is despite the fact that we only considered a 30\% sample size in these experiments; our preliminary results~\cite{Wanye2019FastSampling} suggest that generally a 35\%-45\% sample size is needed to achieve near-optimal accuracy.

For future work, we envision studying the generalizability of SamBaS by applying it to several categories of real-world graphs, including protein interaction networks and social network graphs. Additionally, we would like to see SamBaS applied to overlapping community detection, which is a more accurate representation of the real-world scenario, albeit more computationally expensive.
Finally, while the Partition Quality Score metric works for comparing results on the same graph, it is not suitable for comparing results across different graphs, likely due to the effect that graph size has on $H^{max}$ and $H$. Therefore, to improve the evaluation of community detection algorithms in real-world graphs without known ground truth, we plan to look into different normalization schemes that would make Partition Quality Score comparable across graphs.

\section{Conclusion}

In this manuscript, we present and thoroughly characterize a sampling-based stochastic block partitioning~(SamBaS) approach to accelerate community detection. We generate graphs with community structure based on the degree-corrected stochastic blockmodel~(DCSBM), and we perform a series of experiments to determine the effects of various graph parameters on the quality of results and speedup obtained with SamBaS across different sampling algorithms and different sample sizes. These experiments show that SamBaS is flexible; the choice of sampling algorithm (see 
Appendix~A.1) and sample size (see Figure~\ref{fig:intro}) allows the user to make a trade-off between speedup and accuracy. This is exemplified by SamBaS achieving a speedup of up to 10$\times$ without sacrificing result quality or by improving result quality (i.e., F1 score) by over 150\% at a speedup of around 2$\times$. If some degradation in result quality is not an issue, SamBaS can achieve speedups as high as 45$\times$ (see 
Appendix~A.3). Moreover, the experimental results show that SamBaS improves accuracy on graphs with weak community structure, as characterized by low community strength, maximum vertex degree, and density.


Additionally, we introduce a new metric, \textit{Partition Quality Score}, for evaluating the results of community detection in real-world graphs without known ground-truth community memberships. Typically, this is done using the Modularity metric.
However, we empirically show that Modularity does not correlate well with F1 Scores on our synthetic graphs, while Partition Quality Score does. 
Using this new metric, we show that SamBaS produces results of comparable quality to running stochastic block partitioning without sampling on real-world web graphs while providing significant speedups at the 30\% sample size.



\appendices

\section{Synthetic Graph Results}~\label{appendix:synthetic}

In this section, we present unabridged results on synthetic graphs with all 5 sampling algorithms.

\subsection{Sampling Algorithm Comparison}\label{appendix:result_summary}

\begin{figure}[h!]
  \centering
  \includegraphics[width=\columnwidth]{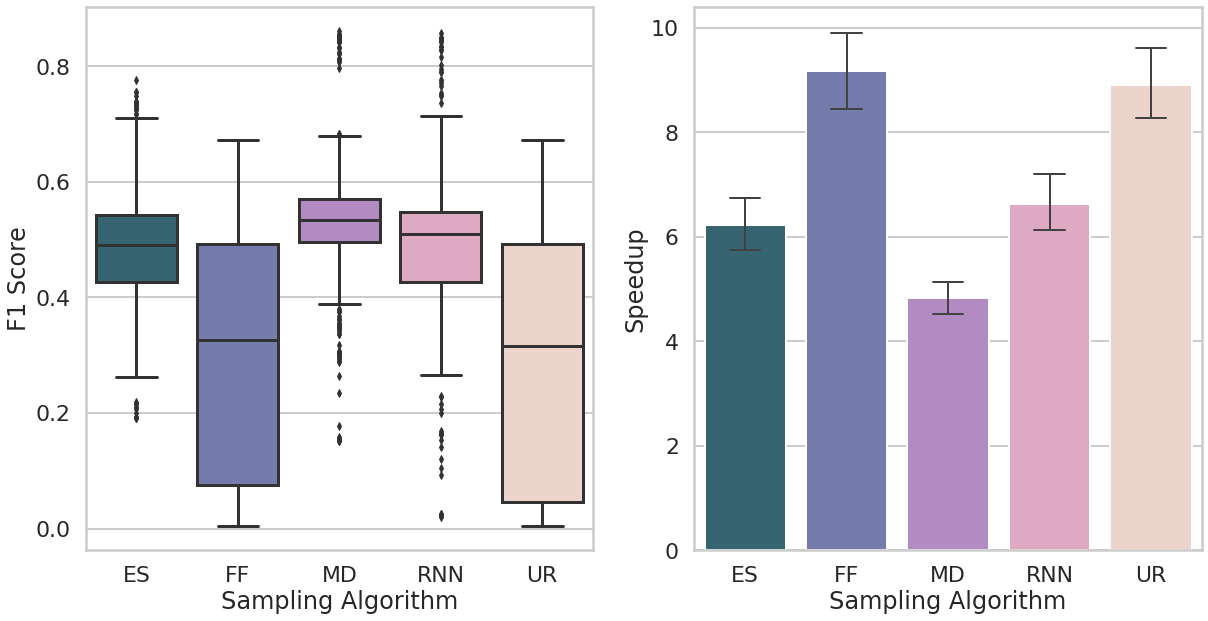}
  \caption{Comparison of F1 scores (left) and average speedups compared to performing community detection without sampling (right) obtained using each of the five sampling algorithms, across all 31 synthetic graphs. Data includes the 10\%, 30\% and 50\% sample sizes.}
  \label{fig:summary}
\end{figure}

Across 31 graphs, we evaluate the performance of all five of our thresholded sampling algorithms.
The results, summarized in Figure~\ref{fig:summary}, show that
MD sampling leads to the highest F1 scores on average, but also the lowest average speedup at 4.8$\times$. FF and UR sampling perform the worst, with the lowest average F1 scores and the highest variability, due to the presence of island vertices in the sampled graph, but lead to the highest average speedups.

\subsection{F1 Score Results}

\begin{table*}[h!]
\centering
\caption{Average F1 Scores for synthetic graph experiments}
\label{tab:appendix_all_f1scores}
\resizebox{0.92\textwidth}{!}{
\begin{tabular}{c|>{\bfseries}c|cc>{\bfseries}ccc|cc>{\bfseries}ccc|cc>{\bfseries}ccc}
\hline
Graph &  & \multicolumn{5}{c|}{50\% Sample Size} & \multicolumn{5}{c|}{30\% Sample Size} & \multicolumn{5}{c}{10\% Sample Size} \\
ID & None & ES & FF & MD & RNN & \normalfont{UR} & ES & FF & MD & RNN & \normalfont{UR} & ES & \normalfont{FF} & MD & RNN & UR \\
\hline
S1 & 0.54 & 0.51 & 0.49 & 0.54 & 0.51 & 0.50 & 0.50 & 0.38 & 0.52 & 0.51 & 0.31 & 0.40 & 0.08 & 0.49 & 0.39 & 0.04 \\
S2 & 0.58 & 0.53 & 0.51 & 0.55 & 0.53 & 0.49 & 0.50 & 0.34 & 0.53 & 0.51 & 0.32 & 0.42 & 0.06 & 0.50 & 0.41 & 0.04 \\
S3 & 0.59 & 0.51 & 0.50 & 0.54 & 0.52 & 0.50 & 0.48 & 0.32 & 0.52 & \textbf{0.52} & 0.30 & 0.44 & 0.05 & 0.51 & 0.43 & 0.03 \\
S4 & 0.62 & 0.55 & 0.50 & 0.56 & 0.54 & 0.51 & 0.49 & 0.28 & 0.55 & 0.54 & 0.27 & 0.40 & 0.04 & 0.51 & 0.41 & 0.02 \\
S5 & 0.59 & 0.56 & 0.51 & 0.57 & 0.56 & 0.49 & 0.49 & 0.28 & 0.54 & 0.53 & 0.26 & 0.40 & 0.04 & 0.51 & 0.44 & 0.02 \\
S6 & 0.60 & 0.57 & 0.48 & 0.58 & 0.57 & 0.48 & 0.50 & 0.21 & 0.54 & 0.53 & 0.23 & 0.39 & 0.03 & 0.51 & 0.44 & 0.02 \\
S7 & 0.55 & 0.67 & 0.65 & 0.68 & 0.66 & 0.65 & \textbf{0.65} & 0.50 & 0.65 & \textbf{0.65} & 0.51 & 0.60 & 0.25 & 0.65 & 0.59 & 0.20 \\
S8 & 0.59 & \textbf{0.62} & 0.58 & 0.62 & 0.60 & 0.57 & 0.58 & 0.43 & 0.60 & \textbf{0.60} & 0.44 & 0.55 & 0.12 & 0.60 & 0.56 & 0.08 \\
S9 & 0.58 & 0.57 & 0.53 & 0.58 & 0.56 & 0.53 & 0.50 & 0.35 & \normalfont{0.54} & \textbf{0.55} & 0.32 & 0.44 & 0.06 & 0.53 & 0.45 & 0.04 \\
S10 & 0.51 & 0.46 & 0.42 & 0.48 & 0.45 & 0.42 & 0.39 & 0.24 & 0.44 & 0.42 & 0.19 & 0.32 & 0.03 & 0.40 & 0.33 & 0.02 \\
S11 & 0.41 & 0.34 & 0.32 & 0.36 & 0.35 & 0.32 & 0.28 & 0.14 & 0.34 & 0.31 & 0.11 & 0.21 & 0.02 & 0.30 & 0.16 & 0.01 \\
S12 & 0.58 & 0.54 & 0.50 & 0.55 & 0.54 & 0.49 & 0.49 & 0.31 & 0.51 & \textbf{0.51} & 0.33 & 0.42 & 0.04 & 0.50 & 0.41 & 0.02 \\
S13 & 0.56 & 0.53 & 0.49 & 0.54 & 0.53 & 0.50 & 0.48 & 0.29 & 0.52 & 0.51 & 0.29 & 0.38 & 0.05 & 0.50 & 0.43 & 0.03 \\
S14 & 0.56 & 0.54 & 0.50 & 0.56 & 0.52 & 0.50 & 0.50 & 0.31 & 0.52 & 0.51 & 0.29 & 0.39 & 0.05 & 0.50 & 0.44 & 0.03 \\
S15 & 0.58 & 0.55 & 0.50 & 0.57 & 0.53 & 0.51 & 0.49 & 0.32 & 0.53 & 0.51 & 0.32 & 0.43 & 0.06 & 0.50 & 0.41 & 0.04 \\
S16 & 0.60 & 0.57 & 0.53 & 0.60 & 0.54 & 0.51 & 0.50 & 0.34 & 0.55 & 0.53 & 0.31 & 0.41 & 0.06 & 0.54 & 0.44 & 0.04 \\
S17 & 0.13 & 0.33 & 0.33 & 0.36 & 0.33 & 0.33 & 0.20 & 0.16 & 0.30 & 0.22 & 0.15 & \textbf{0.31} & 0.01 & \normalfont{0.16} & 0.02 & 0.01 \\
S18 & 0.41 & 0.49 & 0.47 & 0.51 & 0.47 & 0.47 & 0.44 & 0.29 & \normalfont{0.42} & \textbf{0.46} & 0.29 & 0.38 & 0.04 & 0.45 & 0.36 & 0.02 \\
S19 & 0.56 & 0.55 & 0.49 & 0.56 & 0.53 & 0.51 & 0.49 & 0.34 & 0.52 & 0.51 & 0.33 & 0.42 & 0.05 & 0.50 & 0.39 & 0.03 \\
S20 & 0.61 & 0.54 & 0.53 & 0.57 & 0.55 & 0.52 & 0.51 & 0.34 & 0.55 & 0.54 & 0.33 & 0.44 & 0.07 & 0.51 & 0.44 & 0.03 \\
S21 & 0.66 & 0.59 & 0.58 & 0.60 & 0.58 & 0.56 & 0.54 & 0.39 & 0.56 & 0.54 & 0.37 & 0.49 & 0.09 & 0.52 & 0.46 & 0.04 \\
S22 & 0.41 & 0.51 & 0.51 & 0.53 & 0.50 & 0.50 & 0.47 & 0.32 & \normalfont{0.46} & \textbf{0.49} & 0.30 & 0.45 & 0.05 & 0.49 & 0.38 & 0.03 \\
S23 & 0.47 & 0.51 & 0.49 & 0.54 & 0.50 & 0.49 & 0.47 & 0.33 & 0.49 & 0.48 & 0.33 & 0.44 & 0.05 & 0.48 & 0.41 & 0.03 \\
S24 & 0.57 & 0.52 & 0.51 & 0.55 & 0.52 & 0.51 & 0.48 & 0.31 & \normalfont{0.48} & \textbf{0.49} & 0.30 & 0.41 & 0.05 & 0.48 & 0.41 & 0.03 \\
S25 & 0.61 & 0.55 & 0.52 & 0.57 & 0.54 & 0.51 & 0.49 & 0.31 & 0.56 & 0.52 & 0.31 & 0.42 & 0.05 & 0.51 & 0.43 & 0.03 \\
S26 & 0.63 & 0.55 & 0.51 & 0.57 & 0.54 & 0.53 & 0.50 & 0.35 & 0.56 & 0.54 & 0.32 & 0.46 & 0.04 & 0.52 & 0.42 & 0.03 \\
S27 & 0.24 & 0.55 & 0.54 & 0.59 & 0.55 & 0.55 & 0.49 & 0.43 & 0.56 & 0.49 & 0.42 & \textbf{0.28} & 0.14 & \normalfont{0.21} & 0.12 & 0.09 \\
S28 & 0.87 & 0.72 & 0.57 & 0.81 & 0.80 & 0.57 & 0.55 & 0.36 & 0.81 & 0.64 & 0.36 & 0.33 & 0.11 & 0.39 & 0.28 & 0.11 \\
S29 & 0.90 & 0.73 & 0.58 & 0.84 & 0.83 & 0.58 & 0.54 & 0.40 & 0.83 & 0.71 & 0.39 & 0.38 & 0.07 & 0.41 & 0.37 & 0.06 \\
S30 & 0.89 & 0.74 & 0.63 & 0.85 & \textbf{0.85} & 0.62 & 0.57 & 0.46 & 0.86 & 0.75 & 0.46 & 0.44 & 0.08 & 0.49 & 0.44 & 0.07 \\
S31 & 0.91 & 0.76 & 0.67 & 0.85 & 0.84 & 0.66 & 0.62 & 0.53 & 0.85 & 0.77 & 0.53 & 0.48 & 0.05 & 0.56 & 0.52 & 0.06 \\
\hline
\end{tabular}
}
\end{table*}

\begin{table*}[h!]
\centering
\caption{Average speedup for synthetic graph experiments}
\label{tab:appendix_all_speedups}
\resizebox{0.92\textwidth}{!}{
\begin{tabular}{c|rrrr>{\bfseries}r|rrrr>{\bfseries}r|r>{\bfseries}rrrr}
\hline
Graph  & \multicolumn{5}{c|}{50\% Sample Size} & \multicolumn{5}{c|}{30\% Sample Size} & \multicolumn{5}{c}{10\% Sample Size} \\
ID & \multicolumn{1}{c}{ES} & \multicolumn{1}{c}{FF} & \multicolumn{1}{c}{MD} & \multicolumn{1}{c}{RNN} & \multicolumn{1}{c|}{UR} & \multicolumn{1}{c}{ES} & \multicolumn{1}{c}{FF} & \multicolumn{1}{c}{MD} & \multicolumn{1}{c}{RNN} & \multicolumn{1}{c|}{UR} & \multicolumn{1}{c}{ES} & \multicolumn{1}{c}{FF} & \multicolumn{1}{c}{MD} & \multicolumn{1}{c}{RNN} & \multicolumn{1}{c}{UR} \\
\hline
S1 & 1.60 & 2.28 & 1.99 & 2.12 & 2.37 & 4.30 & 5.07 & 3.14 & 3.81 & 5.28 & 10.03 & 11.74 & 8.33 & 10.09 & 11.40 \\
S2 & 1.32 & 2.24 & 1.86 & 2.00 & 2.31 & 4.11 & 5.24 & 3.22 & 4.01 & 5.31 & 10.74 & 13.73 & 9.01 & 11.57 & 13.47 \\
S3 & 1.09 & 2.20 & 1.84 & 2.01 & 2.32 & 3.94 & 5.13 & 3.09 & 3.81 & 5.25 & 11.27 & 15.33 & 9.01 & 11.95 & 13.95 \\
S4 & 0.95 & 2.42 & 1.93 & 2.13 & 2.47 & 4.39 & 6.05 & 3.53 & 4.44 & 6.20 & 11.78 & 17.14 & 9.50 & 13.47 & 16.48 \\
S5 & 0.61 & 2.41 & 1.86 & 2.03 & 2.42 & 3.94 & 5.59 & 3.17 & 4.02 & 5.81 & 10.82 & 17.36 & 8.69 & 12.86 & 14.07 \\
S6 & 0.43 & \textbf{2.55} & 1.94 & 2.10 & \normalfont{2.53} & 3.74 & \textbf{6.13} & 3.07 & 4.02 & \normalfont{5.98} & 11.40 & 19.95 & 9.18 & 14.24 & 16.99 \\
S7 & 1.24 & 2.36 & 1.98 & 2.09 & 2.39 & 3.96 & \textbf{5.74} & 3.21 & 3.93 & \normalfont{5.68} & 12.28 & 20.84 & 9.47 & 13.90 & 19.49 \\
S8 & 1.18 & \textbf{2.44} & 1.97 & 2.21 & 2.44 & 4.25 & 5.77 & 3.33 & 4.16 & 5.78 & 12.47 & 18.60 & 10.04 & 14.45 & 18.22 \\
S9 & 1.05 & 2.27 & 1.86 & 2.04 & 2.36 & 4.00 & 5.39 & 3.16 & 4.08 & 5.46 & 11.50 & 15.51 & 8.94 & 12.71 & 14.87 \\
S10 & 1.03 & 2.38 & 1.95 & 2.13 & 2.44 & 4.27 & 5.44 & 3.35 & 4.20 & 5.52 & 11.17 & 14.56 & 9.25 & 11.95 & 12.92 \\
S11 & 1.16 & 2.55 & 2.04 & 2.34 & 2.68 & 4.76 & 5.86 & 3.63 & 4.66 & 5.94 & 10.85 & 14.21 & 9.46 & 11.19 & 12.33 \\
S12 & 1.10 & 2.38 & 2.00 & 2.08 & 2.49 & 4.40 & 5.61 & 3.42 & 4.32 & 5.67 & 11.99 & 14.95 & 9.07 & 13.67 & 14.65 \\
S13 & 1.09 & 2.40 & 1.97 & 2.15 & 2.52 & 4.38 & \textbf{5.50} & 3.39 & 4.28 & \normalfont{5.49} & 12.33 & 16.59 & 9.35 & 13.08 & 15.46 \\
S14 & 1.10 & 2.37 & 1.93 & 2.10 & 2.43 & 4.28 & 5.51 & 3.32 & 4.20 & 5.61 & 11.79 & 15.86 & 9.33 & 13.32 & 15.41 \\
S15 & 1.01 & 2.23 & 1.88 & 2.03 & 2.35 & 3.97 & 5.27 & 3.17 & 4.02 & 5.42 & 11.71 & 15.73 & 9.24 & 12.79 & 14.63 \\
S16 & 1.10 & 2.30 & 1.90 & 2.06 & 2.37 & 4.29 & 5.82 & 3.40 & 4.25 & 5.89 & 11.94 & 16.77 & 9.03 & 12.76 & 16.42 \\
S17 & 1.20 & \textbf{2.29} & 1.83 & 2.00 & \normalfont{2.26} & 4.31 & \textbf{5.19} & 3.24 & 4.09 & \normalfont{5.16} & 10.52 & 25.31 & 9.02 & 14.78 & 24.70 \\
S18 & 1.20 & 2.43 & 2.01 & 2.17 & 2.57 & 4.48 & \textbf{5.80} & 3.57 & 4.29 & \normalfont{5.59} & 12.44 & 17.15 & 9.70 & 13.76 & 16.47 \\
S19 & 1.09 & 2.39 & 2.04 & 2.15 & 2.50 & 4.47 & \textbf{5.87} & 3.49 & 4.42 & \normalfont{5.82} & 12.47 & 16.87 & 10.17 & 13.50 & 15.66 \\
S20 & 1.07 & 2.44 & 1.97 & 2.14 & 2.47 & 4.28 & 5.56 & 3.43 & 4.30 & 5.75 & 12.13 & 16.02 & 9.80 & 12.99 & 14.85 \\
S21 & 1.00 & 2.25 & 1.89 & 2.04 & 2.33 & 4.15 & 5.40 & 3.38 & 4.14 & 5.47 & 12.14 & 15.28 & 9.95 & 13.34 & 14.20 \\
S22 & 0.98 & 2.29 & 2.04 & 2.15 & 2.36 & 4.39 & \textbf{5.52} & 3.54 & 4.37 & 5.52 & 12.92 & 17.07 & 10.01 & 14.23 & 14.39 \\
S23 & 0.94 & 2.19 & 1.93 & 2.07 & 2.29 & 4.07 & 5.19 & 3.36 & 3.94 & 5.24 & 11.81 & 15.69 & 9.40 & 13.20 & 14.25 \\
S24 & 1.03 & 2.39 & 1.92 & 2.11 & 2.42 & 4.25 & \textbf{5.66} & 3.43 & 4.24 & \normalfont{5.50} & 12.12 & 16.85 & 9.72 & 13.41 & 15.40 \\
S25 & 1.16 & 2.26 & 1.85 & 2.04 & 2.33 & 4.22 & 5.51 & 3.34 & 4.16 & 5.60 & 11.94 & 15.94 & 8.81 & 13.01 & 15.19 \\
S26 & 1.16 & 2.30 & 1.78 & 2.01 & 2.36 & 4.19 & 5.43 & 3.32 & 4.08 & 5.44 & 11.39 & 16.26 & 8.68 & 12.68 & 15.18 \\
S27 & 2.17 & 2.45 & 1.93 & 2.18 & 2.54 & 4.34 & 5.79 & 3.34 & 4.44 & 6.05 & 11.60 & 12.54 & 9.60 & 12.19 & 12.04 \\
S28 & 2.24 & 4.68 & 1.17 & 1.35 & 4.75 & 4.78 & 11.82 & 1.68 & 2.86 & 11.99 & 14.97 & 21.21 & 9.58 & 16.19 & 20.81 \\
S29 & 2.73 & 6.08 & 1.28 & 1.44 & 6.11 & 7.63 & 17.46 & 1.97 & 3.20 & 18.33 & 23.62 & \normalfont{30.91} & 12.09 & 23.73 & \textbf{34.28} \\
S30 & 3.06 & \textbf{5.76} & 1.17 & 1.40 & \normalfont{5.62} & 8.10 & \textbf{19.87} & 1.79 & 2.90 & \normalfont{19.85} & 25.07 & 37.57 & 11.99 & 24.67 & 36.39 \\
S31 & 3.44 & 6.76 & 1.28 & 1.48 & 7.05 & 10.57 & 23.73 & 1.98 & 3.12 & 23.88 & 30.17 & 45.28 & 11.57 & 25.00 & 43.82 \\
\hline
\end{tabular}
}
\end{table*}

Table~\ref{tab:appendix_all_f1scores} shows the average F1 scores obtained for each graph by each sampling algorithm, including the baseline (None). In 24 out of the 31 graphs, the difference between the F1 scores obtained with our sampling approach and the baseline is greater than or equal to $-0.05$. On the graph where our approach performs worst when compared to the baseline, S31, we 
can still obtain an F1 score difference of just $0.06$.

\subsection{Speedup Results}

Table~\ref{tab:appendix_all_speedups} shows the speedups obtained for each graph by each sampling algorithm over the baseline. Uniform Random (UR) and Forest Fire (FF) sampling always achieve the best speedups, due to the fast sampling time and lower density of the resulting sampled graphs. However, the quality of community detection results obtained with these algorithms is low. Of the remaining 3 algorithms, Random Node Neighbor (RNN) sampling results in the best speedups at the 50\% sample sizes, though Expansion Snowball (ES) is faster at the 30\% and 10\% sample sizes.

\section{Web Graph Results}~\label{appendix:real}

In this section, we present unabridged results on real-world graphs with the 3 sampling algorithms found to perform best on synthetic graphs.

\subsection{Quality Score Results}

Table~\ref{tab:appendix_real_quality_scores} shows the average Partition Quality Scores~(PQS) obtained for each graph by each sampling algorithm at the 30\% sample size, including the baseline (None). In 6 out of the 10 graphs, Maximum Degree (MD) sampling results in the highest PQS. However, the PQS on all graphs except R10 is close, suggesting that the solutions obtained with all 3 sampling algorithms are similar in how well they describe the graph. On R10, Expansion Snowball (ES) sampling outperforms the other two sampling algorithms by a wide margin. Note that all the PQS obtained with SamBaS are very close to the baseline, showing very good result quality preservation.

\begin{table}[h!]
\centering
\caption{Average Partition Quality Score for real-world graph experiments.}
\label{tab:appendix_real_quality_scores}
\begin{tabular}{c|>{\bfseries}r|rrr}
\hline
Graph ID & \normalfont{None} & \multicolumn{1}{c}{ES} & \multicolumn{1}{c}{MD} & \multicolumn{1}{c}{RNN} \\
\hline
R1 & 0.67 & \textbf{0.66} & \textbf{0.66} & \textbf{0.66} \\
R2 & 0.68 & 0.65 & \textbf{0.66} & 0.65 \\
R3 & 0.66 & 0.64 & \textbf{0.65} & \textbf{0.65} \\
R4 & 0.72 & 0.69 & 0.70 & \textbf{0.71} \\
R5 & 0.68 & 0.65 & \textbf{0.68} & 0.67 \\
R6 & 0.45 & 0.42 & \textbf{0.45} & 0.44 \\
R7 & 0.56 & 0.55 & \textbf{0.56} & \textbf{0.56} \\
R8 & 0.68 & 0.66 & \textbf{0.68} & 0.66 \\
R9 & 0.70 & 0.67 & \textbf{0.69} & \textbf{0.69} \\
R10 & 0.84 & \textbf{0.83} & 0.54 & 0.61 \\
\hline
\end{tabular}
\end{table}

\subsection{Speedup Results}

\begin{table}[h!]
\centering
\caption{Average speedup for real-world graph experiments.}
\label{tab:appendix_real_speedup}
\begin{tabular}{c|rrr}
\hline
Graph ID & \multicolumn{1}{c}{ES} & \multicolumn{1}{c}{MD} & \multicolumn{1}{c}{RNN} \\
\hline
R1 & 2.50 & 2.45 & \textbf{2.68} \\
R2 & \textbf{2.78} & 2.19 & 2.30 \\
R3 & 2.77 & 2.17 & \textbf{3.69} \\
R4 & 2.49 & 1.73 & \textbf{2.92} \\
R5 & \textbf{3.28} & 1.97 & 2.43 \\
R6 & \textbf{3.47} & 2.91 & 3.25 \\
R7 & 2.78 & 2.97 & \textbf{3.80} \\
R8 & \textbf{4.28} & 1.16 & 1.17 \\
R9 & 3.55 & 2.20 & \textbf{3.60} \\
R10 & \textbf{20.58} & 3.84 & 5.75 \\
\hline
\end{tabular}
\end{table}

Table~\ref{tab:appendix_real_speedup} shows the average speedup obtained on each web graph by each sampling algorithm at the 30\% sample size. The algorithm that produces the highest speedup is always either the Random Node Neighbor (RNN) or the Expansion Snowball (ES) algorithm, which is consistent with the results obtained on synthetic graphs. The greatest speedup obtained is 20.58$\times$ on R10 using ES sampling.



\ifCLASSOPTIONcompsoc
  \section*{Acknowledgments}
\else
  \section*{Acknowledgment}
\fi

Computing resources supporting this work were provided by Intel\textregistered Development Tools through the Intel\textregistered DevCloud.

We thank Atharva Gondhalekar and Sajal Dash for their insightful comments and discussion.

\ifCLASSOPTIONcaptionsoff
  \newpage
\fi

\bibliographystyle{IEEEtran}
\bibliography{IEEEabrv,main}
%



%

\begin{IEEEbiography}{Frank Wanye} received the
B.Sc. degree from Grand Valley State University, Michigan, USA, in 2014. He is currently working towards a Ph.D. degree in computer science at Virginia Tech, Virginia, USA, where he is advised by Dr. Wu-chun Feng. His research interests include parallel and distributed computing, graph analytics, big data, and machine learning.
\end{IEEEbiography}

\begin{IEEEbiography}{Vitaliy Gleyzer}
Vitaliy Gleyzer has been a staff member at MIT Lincoln Laboratory for over ten years. Prior to joining the Laboratory, he received his Master’s degree in Electrical and Computer Engineering from Carnegie Mellon University, with a research concentration on wide area networks. His current work and research interests are primarily focused on hardware and algorithm co-design for development of novel high-performance hardware architectures, large-scale graph analytics and embedded artificial intelligence applications.
\end{IEEEbiography}

\begin{IEEEbiography}{Edward Kao}
Edward Kao is a research scientist in the AI Software Architecture and Algorithms Group at MIT Lincoln Laboratory. Since 2008, he has conducted research in the exploitation of graph and network data, where actionable information is derived from interactions and relationships between entities. He received a PhD in statistics from Harvard with a dissertation on network causal experiments. Other areas of expertise include analysis of social media influence operations, statistical models for networks, high performance computing for large networks, and experimental design / optimal sampling for network inference. With over twenty journal and conference publications, he was the recipient of the R\&D 100 Award, Tom R. Ten Have Research Award, IEEE HPEC Best Paper Award, Google Best Poster Award, National Defense Science \& Engineering Graduate Fellowship, and MIT Lincoln Scholarship.
\end{IEEEbiography}




\begin{IEEEbiography}{Wu-chun Feng}
Dr. Wu-chun Feng — or more simply, "Wu" — is a professor of computer science and electrical \& computer engineering at Virginia Tech, where he directs the Systems, Networking, and Renaissance Grokking (SyNeRGy) Laboratory. His research interests span many areas of high-performance networking and computing from hardware to applications software.

To the computer science and engineering community, he is perhaps best known for his systems-level research in high-performance networking, ranging from systems-area network architectures such as Quadrics and 10-Gigabit Ethernet (10GigE) to wide-area network frameworks and implementations in support of distributed computing such as adaptive flow control for TCP (i.e., DRS: Dynamic Right-Sizing) and hybrid circuit- and packet-switched networks (i.e., CHEETAH: Circuit-switched High-speed End-to-End Transport ArcHitecture) and the autonomic rate-adaptive protocols that run on them.

To the general scientific community, he is oftentimes referred to as "Mr. Green Destiny" or "The Green Destiny Guy." Green Destiny debuted in early 2002 as the first major instantiation of the Supercomputing in Small Spaces project. It was a 240-processor supercomputer with a footprint of five square feet and a power envelope of a mere 3.2 kilowatts that debuted in early 2002. This supercomputer, which produced an admirable Linpack rating of 101 Gflops, operated without any unscheduled downtime for its two-year lifetime while running in an 85° F warehouse at 7,400 feet above sea level with no air conditioning, no air humidification, and no air filtration. Green Destiny garnered international attention in over 100 media outlets including BBC News, CNN, and The New York Times and led in part to Dr. Feng being named to HPCwire's Top People to Watch List in 2004.

Dr. Feng received a B.S. in Electrical \& Computer Engineering and Music (Honors) and an M.S. in Computer Engineering from Penn State University in 1988 and 1990, respectively. He earned a Ph.D. in Computer Science from the University of Illinois at Urbana-Champaign in 1996. His previous professional stints include IBM T.J. Watson Research Center, NASA Ames Research Center, Vosaic, University of Illinois at Urbana-Champaign, Purdue University, The Ohio State University, Orion Multisystems, and Los Alamos National Laboratory.
\end{IEEEbiography}




\end{document}